\documentclass[review]{elsarticle}

\usepackage{lineno,hyperref}

\journal{ArXiv}

\usepackage{graphicx}
\usepackage{caption}
\usepackage{subcaption}
\usepackage{booktabs}

\begin{document}

\begin{frontmatter}

\title{Forecasting Covid-19 dynamics in Brazil: a data driven approach\tnoteref{mytitlenote}}
\tnotetext[mytitlenote]{This work is supported by CAPES under Grant 001 and by CNPq.}

\newcommand{\orcidauthorA}{0000-0001-7539-4663} 
\newcommand{\orcidauthorB}{0000-0002-8048-8960} 
\newcommand{\orcidauthorC}{0000-0003-0579-8464} 
\newcommand{\orcidauthorD}{0000-0000-0000-000X} 
\newcommand{\orcidauthorE}{0000-0002-7735-5630} 

\author[myprimaryaddress]{Igor G. Pereira}
\author[myprimaryaddress]{Joris M. Guerin}
\author[myprimaryaddress,mysecondaryaddress]{Andouglas G. Silva J\'unior}
\author[mytertiaryaddress]{Cosimo Distante}
\author[myquaternaryaddress]{Gabriel S. Garcia}
\author[myprimaryaddress]{Luiz M. G. Gon\c{c}alves\corref{mycorrespondingauthor}}
\cortext[mycorrespondingauthor]{Corresponding author}
\ead{lmarcos@dca.ufrn.br}

\address[myprimaryaddress]{Federal University of Rio Grande do Norte, Brazil}
\address[mysecondaryaddress]{Federal Institute of Rio Grande do Norte, Brazil}
\address[mytertiaryaddress]{Institute of Applied Sciences and Intelligent Systems, Italy}
\address[myquaternaryaddress]{University of Bras\'ilia, Brazil}

\begin{abstract}
This paper has a twofold contribution. The first is a data driven approach for predicting the Covid-19 pandemic dynamics, based on data from more advanced countries. The second is to report and discuss the results obtained with this approach for Brazilian states, as of May 4th, 2020. We start by presenting preliminary results obtained by training an LSTM-SAE network, which are somewhat disappointing. Then, our main approach consists in an initial clustering of the world regions for which data is available and where the pandemic is at an advanced stage, based on a set of manually engineered features representing a country's response to the early spread of the pandemic. A Modified Auto-Encoder network is then trained from these clusters and learns to predict future data for Brazilian states. These predictions are used to estimate important statistics about the disease, such as peaks. Finally, curve fitting is carried out on the predictions in order to find the distribution that best fits the outputs of the MAE, and to refine the estimates of the peaks of the pandemic. Results indicate that the pandemic is still growing in Brazil, with most states peaks of infection estimated between the 25th of April and the 19th of May 2020. Predicted numbers reach a total of 240 thousand infected Brazilians, distributed among the different states, with S\~ao Paulo leading with almost 65 thousand estimated, confirmed cases. The estimated end of the pandemics (with 97\% of cases reaching an outcome) starts as of May 28th for some states and rests through August 14th, 2020.
\end{abstract}

\begin{keyword}
\texttt{Time Series Prediction}\sep Covid-19 Pandemic \sep Modified Auto-Encoder
\end{keyword}

\end{frontmatter}


\section{Introduction}

The world population is being rapidly infected by the SARS-COV-2 virus pandemic, also known as Covid-19 \cite{Byass2020}. This virus has spread very fast throughout the countries, with a very high contagion rate, reaching all continents in just over three months since the first confirmed case in Asia \cite{Binti2020}. The numbers have grown exponentially reaching, approximately, more than 3.7 million cases and more than a quarter of a million deaths. Distancing and social isolation rules have been used as the only alternative in order to contain the progress of the disease. The flattening of the virus spreading curve that can be modeled according to several approaches \cite{FANELLI2020, Webb2020, Grant2020, Piccolomiini2020, Baerwolff2020} is the first goal strictly related to the rules mentioned. If this does not happen, the number of deaths would skyrocket, as it was recently experimented in countries such as the US, Italy, Spain, France, and the UK. Several warnings about this have been spread in the literature, for example, in the beginning of March 2020,  Fanelli \cite{FANELLI2020} explained that: "In Italy and in other countries that will be facing the epidemic surge soon, this is quite possibly only achievable through a cooperative and disciplined effort of the population as a whole". Successful example of curve flattening have already been seen in Portugal, Germany and South Korea, among others. Other countries, such as New Zealand, managed to limit the spread by completely closing their borders and imposing a complete lock-down to their people. In these countries, the counter-measures provide a certain breath to governments in order to allow the health systems of each country to meet local needs, or to be able to wait for other solutions, such as the development of vaccines (not yet existing so far for the SARS-CoV-2 virus). The medical solutions to minimize or stop the pandemic include in-silico analysis of the SARS-CoV-2 genome \cite{Periwal2020}, aiming to study its weaknesses in an attempt to better understanding it and to develop treatments and vaccines.

In parallel, studies aiming to determine the virus dynamics, its geographical distribution and the peaks of the pandemic in given regions are necessary and useful for correct planning of immediate actions by states management. In this direction, several attempts to model the spread of the virus have been conducted recently \cite{Baerwolff2020, Piccolomiini2020, Grant2020, Distante2020A, WANG2020, Vrugt2020}, including machine learning approaches \cite{Ardabili2020}. Nonetheless, based on reports of these works that were more closely analyzed \cite{Grant2020, Baerwolff2020, Piccolomiini2020, Ardabili2020}, we found that many of these methods rely on parameters that are dependent on the advance of the spreading and on the regional context.

Nonetheless, this is a problem for which time series data are partially available, and using artificial intelligence seems to be a good strategy to devise a method that can predict future data, guided only by past and current data. We argue that a data driven technique can be used to infer the pandemic dynamics from raw data, including future events such as the date of the peak, number of cases and deaths, and the end of the pandemic. Based on this initial assumption, this work tests and validates the possibility of using artificial intelligence tools, mainly the ones based on deep learning, to create models of dissemination and to predict the numbers in different regions, which can be used alternatively or in addition to traditional models \cite{Grant2020, Baerwolff2020, Piccolomiini2020}. We discuss our results in comparison to those obtained with traditional methods and found that they work at least with the same precision in predicting the pandemics events. A preliminary study was carried out in Italy, based on China data with satisfactory results, but with space for improvement \cite{Distante2020A, Distante2020}. An approach based on LSTM has been initially tested by our team in Brazil, and demonstrated problems that are discussed in Section~\ref{sec:lstm_methods}.

So, as the main contribution of this work, we present a way to train a Modified Auto-Encoder (MAE) to forecast virus spreading. The MAE, which demonstrated better results than the preliminary LSTM approach, is chosen and reported as our working model for predicting data from Brazilian states, working at www.natalnet.br/covid. In order to overcome problems as data non-linearity and lack of data due to under notification, we propose an initial clustering of the different countries data, based on Early Mortality, Days until 10x, and Early Acceleration features. Then, different prediction networks are trained within each cluster, using countries that have a more advanced stage of the pandemic than Brazil, e.g. China, Italy, Spain, and the United States. Each networks is then used to make predictions on the Brazilian states belonging to the cluster on which it was trained. We do not have a way to do comparison with traditional approaches, because the pandemics is still in a growing situation in Brazil. However, some results are shown and discussion about the performances are proposed.

Based on the results reported in this paper, up to date, we could verify the applicability of data driven methods to model the Covid-19 dynamics. With this approach, which deals with regional aspects 
of the pandemic, city managers can get more precise information to help then plan their actions. Complementary data about peak prediction and estimated numbers have shown the applicability of our approach to Brazilian states with success. We underline that the findings reported in this paper come from estimated data and cannot be completely guaranteed as being the final truth. However, they are important because they allow managers and even some region population, to have an idea about what the future holds for the pandemic dynamics. We hope that, using the prior expectations of the pandemic curve presented in this paper, better decisions can be taken to help protect the populations.

\section{Materials and Methods}

This work is devoted to develop a method to predict the dynamics of transmission of viral epidemics by analyzing contamination data from the perspective of artificial intelligence. Deep Learning techniques are studied and implemented, aiming to learn the dynamics of the pandemics using data from other locations (countries). This approach is then applied to the specific case of Brazil. We start by describing traditional approaches to set a baseline for comparison, and then detail the different components of the data driven method retained.

\subsection{Modeling virus dynamics (traditional approaches)}

The spread and contamination of the Covid-19 virus is not entirely random and follows certain patterns. These dynamics can vary across different regions as they depend on parameters such as pollution, demographic density, average age of the population, among others. Analyzing the actions taken to fight the virus, in both the social and economic spheres, there is a need for more realistic epidemiological data. Indeed, the use of local models, taking into account the reality of each region, state or municipality, can allow the authorities to take coherent decisions. Therefore, it is assumed that the spread of the virus follows some statistical model, which parameters can be tuned to represent different situations.

Approaches to model the behavior of infectious diseases, such as SEIR, have been used to the epidemic of COVID-19 \cite{Grant2020, Yang2020}. In these approaches, the phase transitions of the disease are modeled as instantaneous rates in differential equations or as probabilities of transition in discrete time differences or matrix equations. These models provide accurate estimates of the position of the equilibrium points, when the rate at which individuals enter each stage is equal to the rate at which they exit. However, they do not accurately capture the distribution of the time an individual spends at each stage; therefore, they do not accurately capture the transitory dynamics of epidemics. Actually, the SEIR model has been tested at Italy \cite{Piccolomiini2020} to model the dynamics of the COVID-19 epidemic. It has been shown to underestimate peak infection rates (by a factor of three using published parameter estimates based on the progress of the epidemic in Wuhan) and to substantially overestimate the persistence of the epidemic after the peak has passed\cite{Grant2020}.

Other approaches such as SIR \cite{RODA2020271}, SEIRD \cite{Piccolomiini2020}, and SEITR \cite{OTUNUGA2020} are also helpful to understanding the Covid-19 dynamics. Nonetheless, the lack of ground truth data prevents us from determining which of these models is the most precise. Despite somehow representing the Covid-19 dynamics, some of these traditional models (SIR, SEIR, SEITR, SEIRD) must be improved so that they can be applied with higher precision to the study of the new virus, as they have been shown to present some issues on the recent works cited above. In this work, besides discussing the main advances of the contributions in this direction, these traditional models are compared to ours, which is a data driven approach. Some preliminary studies on the above methods have been conducted for better understanding of the Covid-19 dynamics. In fact, we verified that it is a virus that cannot be model perfectly with any specific traditional model because of the influence of several factors on its dissemination speed. Mainly, it is difficult to model its behavior because of the non-linearity of infection data caused by under-notifications and also the lack of effective and constant counter measures, which changes all the time as the infection spreads.  For these reasons, it seems appealing to apply AI-based methods. As a first test, we start by implementing an LSTM, one of the default neural network models for analyzing time series data, in the next section.

\subsection{Long Short Term Memory for data training (LSTM)}
\label{sec:lstm_methods}

Several neural network models can be used to solve problems of time series estimation. Recurrent neural networks (RNN) are a family of architectures containing recurring feedback connections, which define an internal state, or short-term memory. This memory makes them suitable for modeling sequential or time series data \cite{Sagheer2020}. To this end, a standard RNN keeps a vector of activation parameters at each time step, especially when short-term dependencies are included in the input data. However, when trained with gradient descent algorithms, learning the long-term dependencies that are encoded in data becomes difficult due to the vanishing gradient problem. This is solved using a specialized neuron for long-term memory that keeps a constant reverse flow in the error signal, allowing it to learn long-term dependencies. This approach was presented by Hochreiter \cite{hochreiter1997long} and is known as LSTM (Long Short Term Memory).

\begin{figure}
    \centering
    \begin{subfigure}[b]{\textwidth}
        \includegraphics[width=\textwidth]{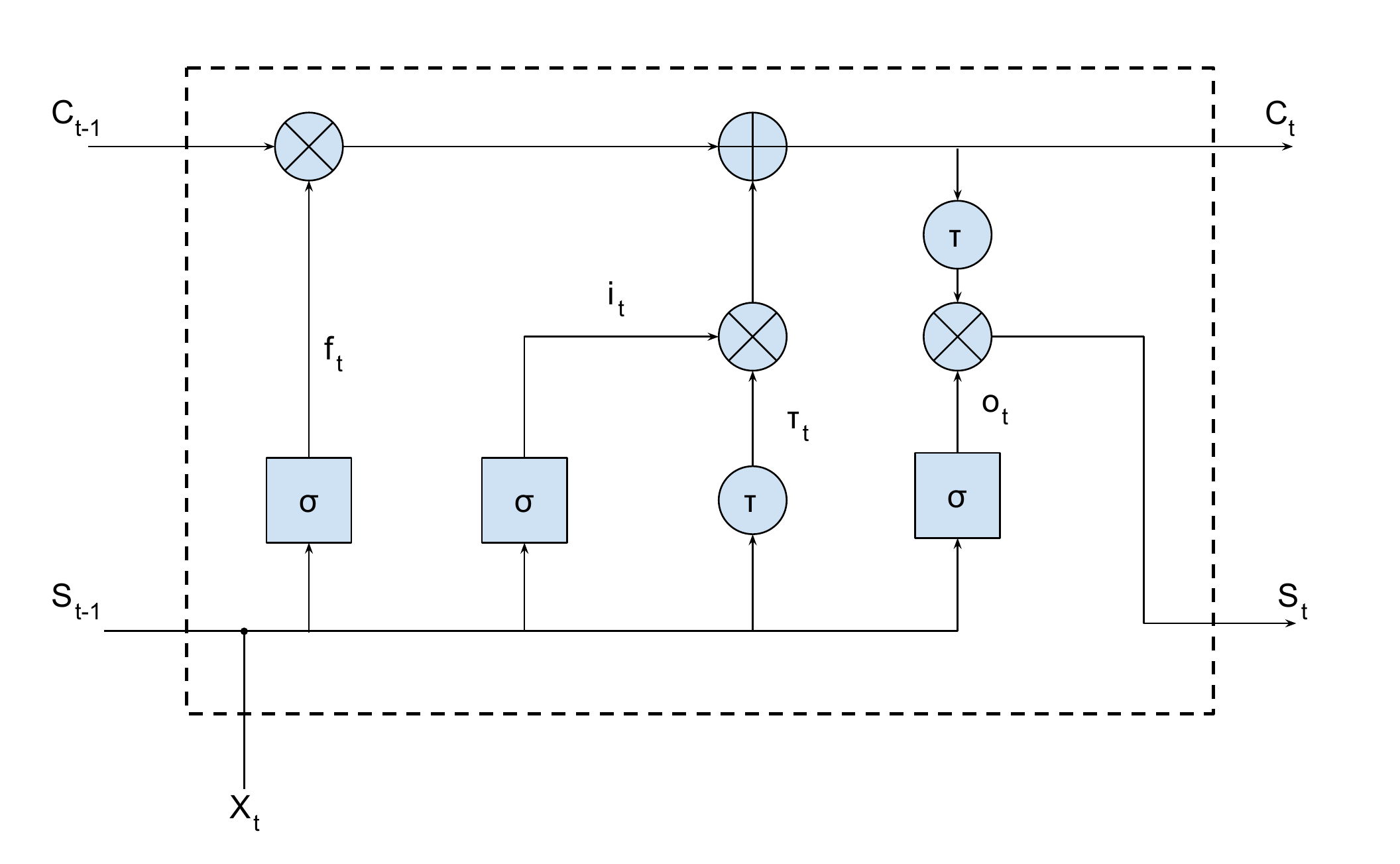}
        \caption{}
        \label{fig:lstm_blocka}
    \end{subfigure}
        \begin{subfigure}[b]{\textwidth}
        \begin{subfigure}[b]{0.3\textwidth}
            \includegraphics[width=\textwidth]{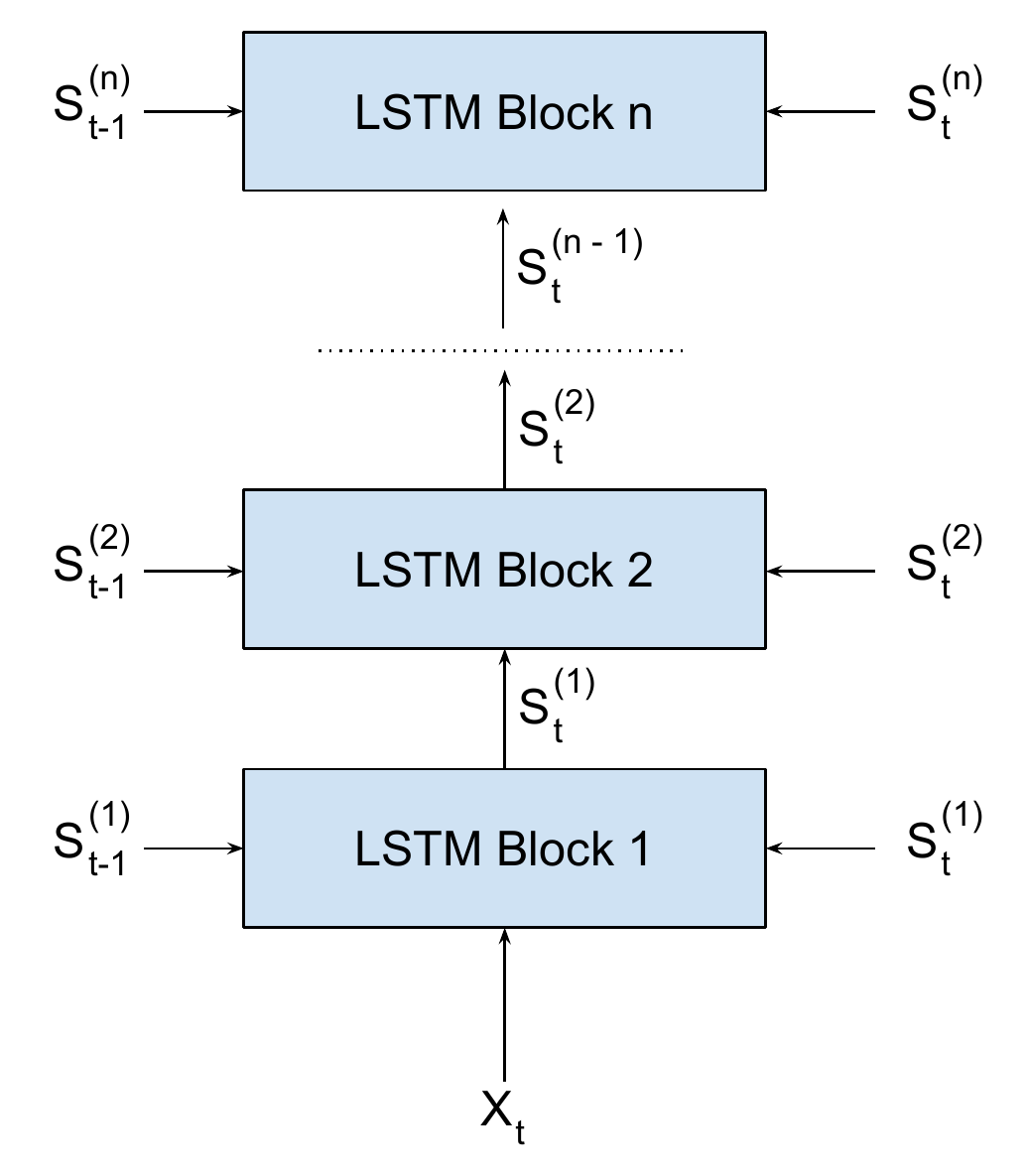}
            \caption{}
            \label{fig:lstm_blockb}
        \end{subfigure}
        \begin{subfigure}[b]{0.6\textwidth}
            \includegraphics[width=\textwidth]{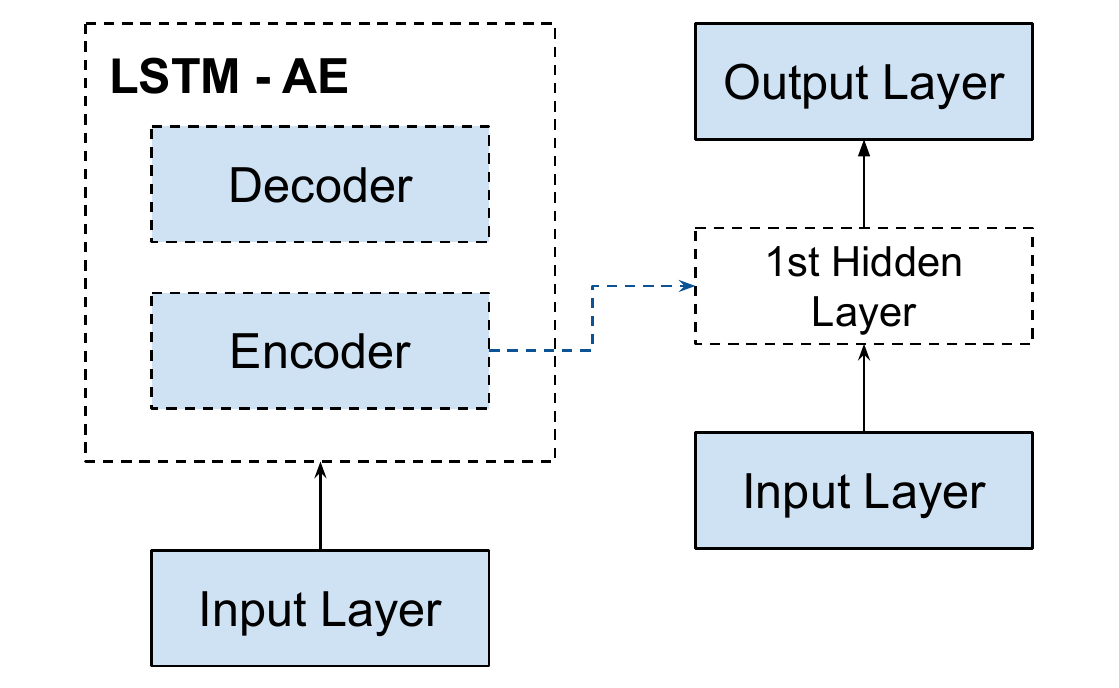}
            \caption{}
            \label{fig:lstm_blockc}
        \end{subfigure}
    \end{subfigure}
    \caption{LSTM, DLSTM and LSTM-SAE Blocks}
    \label{fig:lstm_dlstm_blocks}
\end{figure}

In this way, a LSTM network is kind of RNN architecture, having a recursive branch for modeling time series and solving the vanishing gradient problem. To do so, it uses a memory cell that is able to represent long-term dependencies in the time series, composed of four neural units: input, output, forgetting and the self-recurring neuron (Figure \ref{fig:lstm_blocka}). These units are responsible for controlling the interactions between different memory units. Specifically, the input unit controls whether the input data can modify the state of the memory cell or not. On the other hand, the output unit controls whether or not it can change the state of other memory cells.

Mathematically, considering the output gates ($f_t$, $i_t$, $o_t$ and $\tau_t$) shown in Figure~\ref{fig:lstm_blocka}, we have:

\begin{equation}
    f_t = \sigma(X_t U^f + S^{t-1}W^f + b_f)
\end{equation}
\begin{equation}
    i_t = \sigma(X_t U^i + S^{t-1}W^i + b_i)
\end{equation}
\begin{equation}
    O_t = \sigma(X_t U^o + S^{t-1}W^o + b_o)
\end{equation}
\begin{equation}
    \tau_t = tanh(X_t U^c + S_(t-1) W^c + b_c)
\end{equation}
\begin{equation}
    C_t = C_{t-1} \otimes f_t \oplus i_t \otimes C_ t'
\end{equation}
\begin{equation}
    S_t = O-t \otimes tanh(c_t)
\end{equation}

\noindent where, \textbf{U}, \textbf{W} and \textbf{b} are respectively the input weights, recurrent weights and biases; \textbf{X} is the input; \textbf{S} is the hidden output; \textbf{C} is the cell state; and \textbf{t} is the time step. 

{}

According to Sagheer \cite{Sagheer2020}, despite the advantages of the LSTM architecture, its performance for time series problems is not always satisfactory. The shallow LSTM architecture may not represent the complex features of sequential data efficiently, especially if they are used to learn data from long-range time series with high non-linearity, which is the case for Covid-19 data. In order to overcome this problem, other RNN architectures based on LSTM have been created. We tested two approaches proposed by Sagheer: DLSTM \cite{SAGHEER2019203} and LSTM-SAE \cite{Sagheer2020}, using Covid-19 data from China provinces (daily number of cases and cumulative number of cases).

The LSTM-SAE and DLSTM blocks are shown in Figures~\ref{fig:lstm_blockb} and \ref{fig:lstm_blockc}, respectively. Basically, both blocks are composed of stacked LSTM layers, which increase the depth of the network. Besides that, the LSTM-SAE configuration uses an auto-encoder to initialize the weights of each LSTM layer. In our application, we used only one hidden layer for this setup, but it is possible to use more layers and more auto-encoders as shown on the original paper. In order to select the best architecture for the Covid-19 problem, we trained three models, one LSTM, one DLSTM and one LSTM-SAE. These models were trained using data from all China provinces except Hubei (that was used for testing). We evaluated which model generalized best to the dataset available using the MAPE metric. Finally, we used a dynamic prediction, where the model is updated for each new predicted value. This method improves the forecast due the incorporation of data from other countries or regions. The training parameters and results metrics are shown in Table~\ref{tab:lstms_train_summary}.

\begin{table}[]
\resizebox{\textwidth}{!}{%
\begin{tabular}{@{}lcccccc|cc@{}}
\cmidrule(l){2-9}
\multicolumn{1}{c}{\textbf{}} & \multicolumn{6}{c}{\textbf{Parameters}}                    & \multicolumn{2}{c}{\textbf{Metrics}} \\ \cmidrule(l){2-9} 
 &
  \textbf{Hidden Layers} &
  \textbf{Epochs} &
  \textbf{Epochs AE Model} &
  \textbf{Dropout} &
  \textbf{Units} &
  \textbf{Sequence Lenght} &
  \textbf{MAPE} &
  \textbf{Corelation} \\ \midrule
\textbf{LSTM}                 & 1 & 15 & -  & 0.3 & 4                                  & 5 & 211              & 0.732              \\
\textbf{DLSTM}                & 3 & 15 & -  & 0.3 & \multicolumn{1}{l}{{[}10, 8, 6{]}} & 5 & 92               & 0.798              \\
\textbf{LSTM-SAE}             & 1 & 50 & 15 & 0.3 & 4                                  & 5 & 84               & 0.822              \\ \bottomrule
\end{tabular}%
}
\caption{Training parameters and Metrics}
\label{tab:lstms_train_summary}
\end{table}

\begin{figure}
\centering
\begin{subfigure}[b]{0.49\textwidth}
    \includegraphics[width=\textwidth]{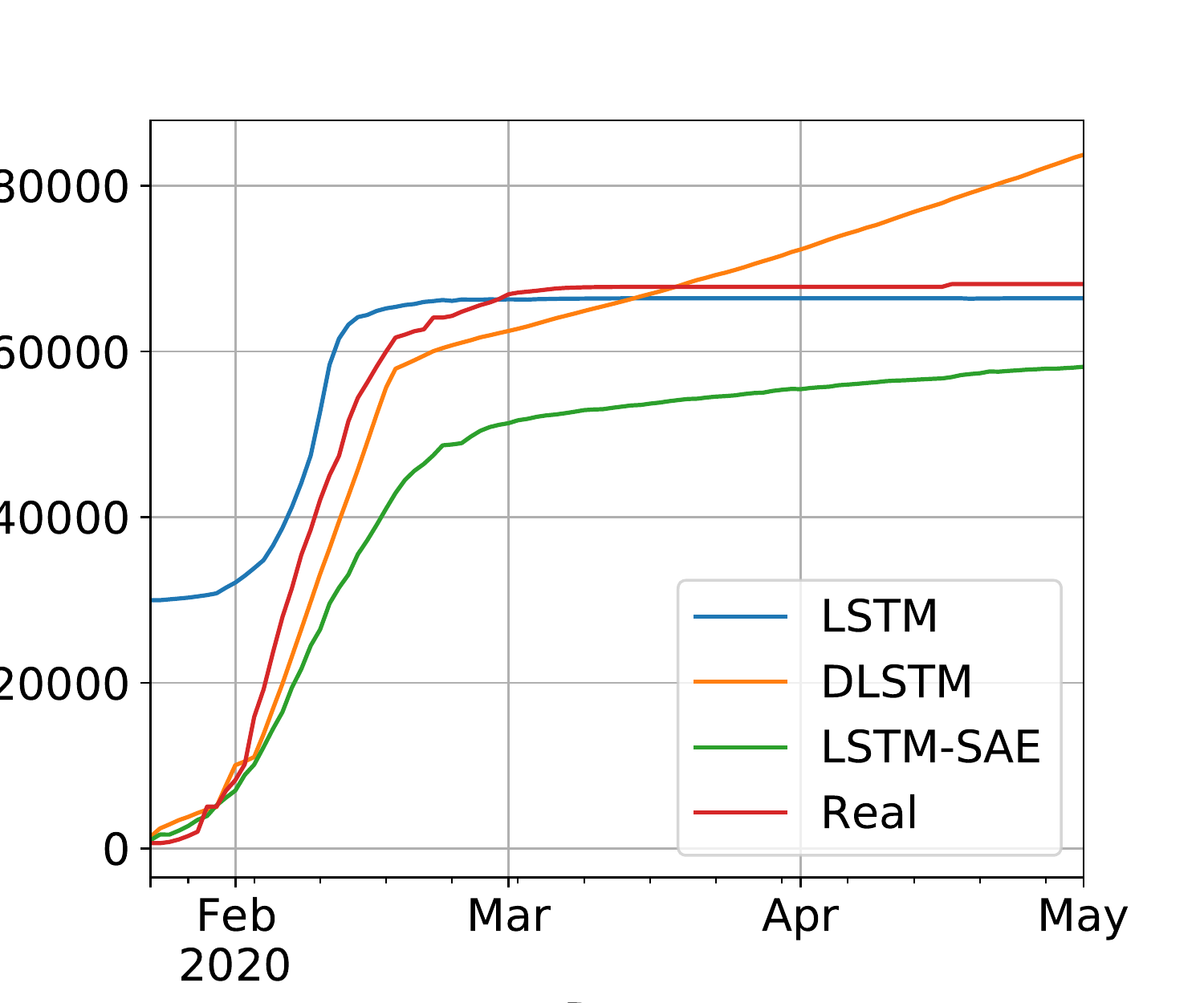}
    \caption{}
    \label{fig:comparison_total}
\end{subfigure}
\begin{subfigure}[b]{0.49\textwidth}
    \includegraphics[width=\textwidth]{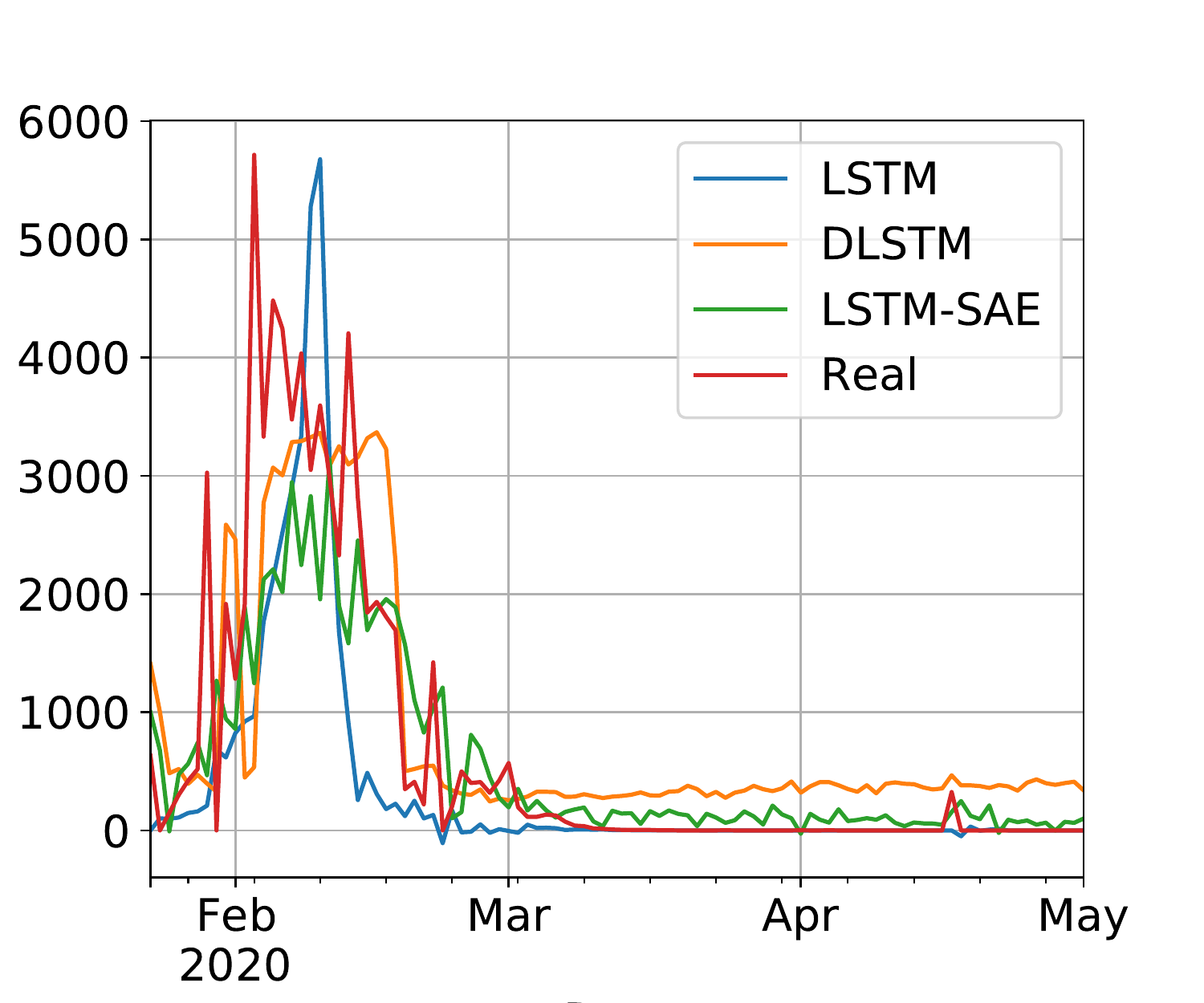}
    \caption{}
    \label{fig:comparison_daily}
\end{subfigure}
\caption{Results for comparison of LSTM, DLSTM and LSTM-SAE on Covid-19 cumulative (\ref{fig:comparison_total}) and daily (\ref{fig:comparison_daily}) number of cases, data from Hubei, province of China}
\label{fig:lstm_dlstm_results}
\end{figure}

Figures~\ref{fig:comparison_total} and \ref{fig:comparison_daily} show the results for the three trained models for Hubei (province of China). As shown in Table~\ref{tab:lstms_train_summary}, the best model was LSTM-SAE, being thus chosen as the model to forecast other regions or countries.

{}

On the one hand, despite the devastating effects of the pandemic, three months of data is a relatively short period of time for training complex time series prediction models without overfitting, which has been reported as one of the main problems for training LSTMs (see Section~\ref{sec:discussion}). On the other hand, this pandemic is the first large scale global pandemic that our generation has to face and there are not yet standardized guidelines for countries on how to react to such an event. For this reason, responses to the pandemic have varied widely throughout the different regions and countries worldwide, thus creating a huge variability in the available data. Hence, we propose to conduct a preliminary study which consists in grouping countries and regions with similar early responses. In this way, smaller specialized networks can be trained for each cluster, and we hope that, by learning on more consistent data, our models could generalize better without overfitting to the training data. Also, we found a better model (MAE) that is used, instead of LSTM, on data resulting from the clustering approach that is described next.

\subsection{Preliminary clustering: Brazilian states in the global context}
\label{sec:methods_clustering}

The objective of this paper is to train a predictive model for Brazil, as well as some distinct models for each of the groups of Brazilian states. Hence, the proposed clustering pipeline considers both entire countries and smaller regions as entries. The input data used in this preliminary study are all countries available in the JHU dataset~\cite{jhu_dataset}, Chinese and Canadian provinces, American, Australian and Brazilian states\cite{brazil_dataset} as well as Italian Regions\cite{italia_dataset}. 

\begin{figure}
    \centering
    \includegraphics[width=\textwidth]{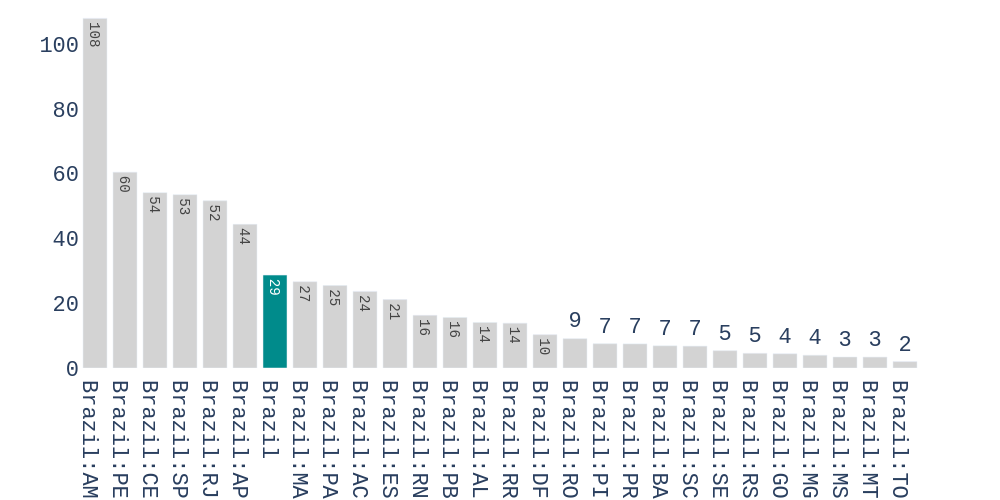}
    \caption{Number of deaths per million inhabitants in the different Brazilian states on 1st of May, 2020}
    \label{fig:DPM_BR}
\end{figure}

The approach used for identifying which countries present similar early responses to Covid-19 is inspired by the literature in this area~\cite{clustering_method}. First, we define the \emph{outbreak date} of a country to be the day at which it registered 5 confirmed cases per million inhabitants. Normalizing by the population of the region helps to characterize the true response of a country, avoiding to give more weight to highly populated countries. Figure~\ref{fig:DPM_BR} shows the number of accumulated deaths per million inhabitants for the different Brazilian states on the first of May, 2020. 

{}

\begin{figure}
\centering
	\centering
	\begin{subfigure}{\textwidth}
	\centering
	\includegraphics[width=\textwidth,height=0.25\textheight]{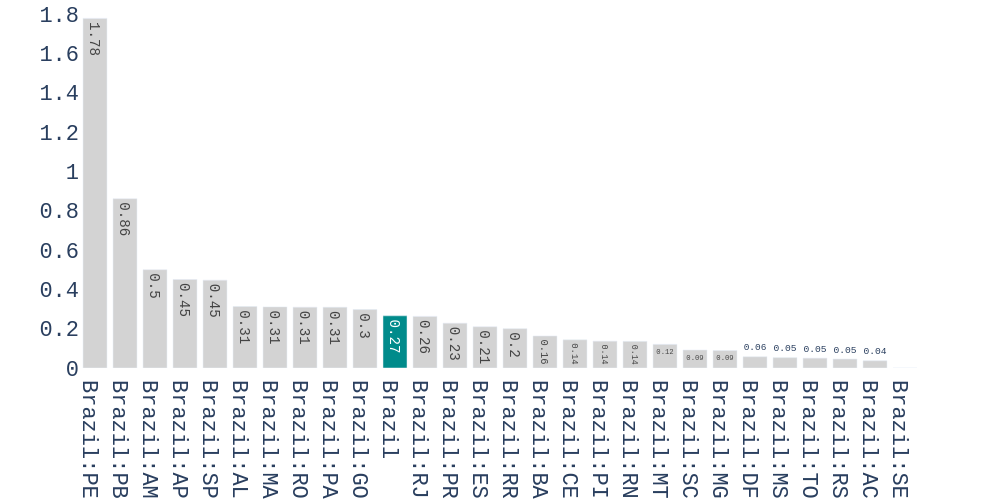}
    \caption{Early Mortality}
	\end{subfigure}
	
	\begin{subfigure}{\textwidth}
	\centering
	\includegraphics[width=\textwidth,height=0.25\textheight]{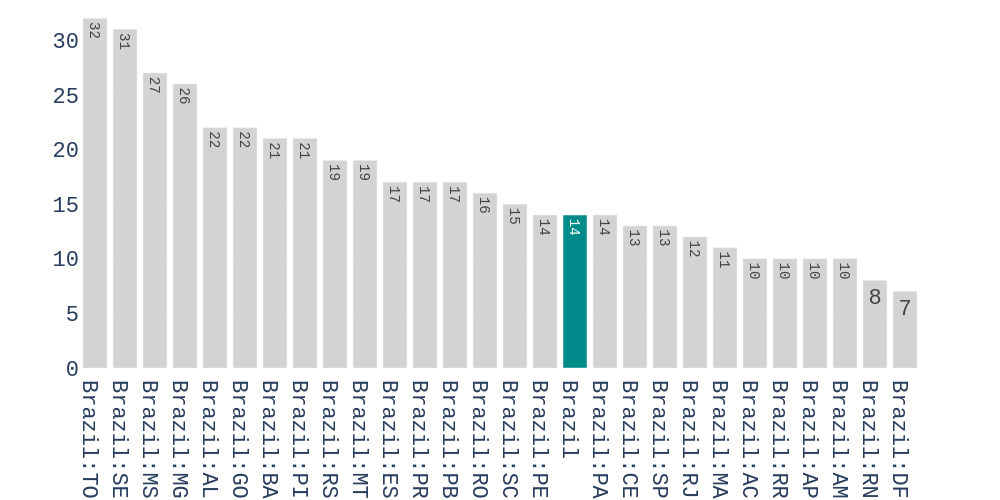}
	\caption{Time to 10x} 
	\end{subfigure}
	
	\begin{subfigure}{\textwidth}
	\centering
	\includegraphics[width=\textwidth,height=0.25\textheight]{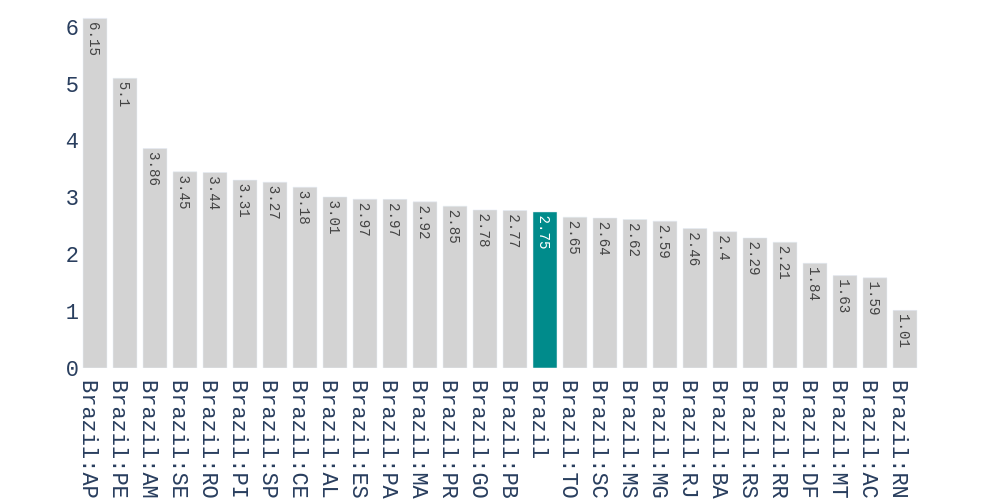}
	\caption{Early Acceleration} 
	\end{subfigure}
	\caption{Values of the three features used for characterizing the early response to covid-19 for the Brazilian states.}
\label{fig:features_BR}
\end{figure}

We start with the preprocessing scheme to be applied on this dataset. A 7-days arithmetic moving average is first calculated to each time series of the dataset. This is done to deal with the seasonality that is observed in data, i.e. higher variability during the weekends. After filtering, a feature representation containing three characteristics is computed for each time series. These features are:
\begin{itemize}
\item \emph{Early Mortality}: Weekly number of deaths 14 days after the outbreak, divided by the number of confirmed cases, in the week of the outbreak. A two weeks period was used because it is the time required to know the outcome of a contamination.
\item \emph{Days until 10x}: The number of days it takes to multiply the confirmed cases by 10, from the day of the outbreak.
\item \emph{Early Acceleration}: If we denote $\Delta_{W_0W_1}$ as the percentage increase of confirmed cases from the week of the outbreak to the week after, and $\Delta_{W_1W_2}$ as he percentage increase from the 1st to the 2nd week after the outbreak, then the early acceleration is defined by: 
\begin{equation}
    earlyAccel = \Delta_{W_1W_2} / \Delta_{W_0W_1}.
\end{equation}{}
\end{itemize}{}
The values of these features for the different Brazilian states are shown on Figure~\ref{fig:features_BR}.
 
{}

Then, the clustering pipeline is applied to the former feature representation to group the different countries/regions together. To do that, a Uniform Manifold Approximation (UMAP) embedding \cite{umap} is applied to generate a two-dimensional clustering friendly feature space. UMAP is an unsupervised embedding method that tends to preserve the global distances present in the initial dataset. This lower dimensional feature space not only facilitates the visualization and interpretation of data but also tends to improve clustering results for algorithms where the number of clusters is unspecified. In practice, UMAP is used with $n\_neighbors = 15$ and $min\_dist = 0$. However, UMAP only produces a new embedded space and does not generates directly the clusters assignments, which are needed to select the countries for training our neural network models. 

To solve this issue, we use the scikit-learn~\cite{scikit-learn} implementation of Affinity Propagation~\cite{affinityPropagation} with a damping factor of $0.8$, applied to the UMAP embedded space. The results from this preliminary clustering procedure are further presented in Section~\ref{sec:results_clustering}. 

Therefore, our clustered data series is ready for the MAE training and prediction procedures, depending on the phase. In practice, to forecast contamination data of a given Brazilian state, we use the time series data of the countries/regions belonging to the same cluster, and which are at a more advanced stage of the pandemic. In this section, the clustering approach adopted to characterize the responses of the different countries is explained, the details of the training process are explained in the following section.

\subsection{Modeling Time-Series with Modified Auto-Encoders}
\label{sec:MAE}

In order to model the transmission dynamics of the SARS-COV2 virus in Brazil, we propose to use a set of Modified Auto-Encoders (MAE) to forecast time-series data regarding the number of daily confirmed cases of Covid-19. An auto-encoder is a specific neural network architecture that is trained to copy its input to its output \cite{Goodfellow-et-al-2016}. In this way, the auto-encoder generates a hidden representation that describes useful properties of the input data. 

The network architecture can be divided in two parts: an encoder function $h=f(x)$, that maps the input data $x$ to the hidden representation $h$, and a decoder function $\hat{x}=g(h)$ that attempts to approximate the input $\hat{x}$ from the hidden representation. With the use of the stochastic gradient descent strategy to train neural network architectures, the auto-encoder mapping functions can be generalized to stochastic mappings such as $p_{encoder}(h|x)$ and $p_{decoder}(\hat{x}|h)$.

The hidden representation, also called latent space, generated by the mapping $p_{encoder}(h|x)$ contains a stochastic representation of the probability distribution of the input data and can be used for dimensionality reduction \cite{Goodfellow-et-al-2016}, feature learning \cite{Goodfellow-et-al-2016}, and also in generative models when combined with latent variable models\cite{Vae2019}.

\subsubsection{The Modified Auto-Encoder proposal}

Auto-encoders can also learn useful properties from time-series if a sequence is applied to its inputs. Such properties may be used to forecast next samples of the given input sequence. In this way, we propose to modify the traditional auto-encoder architecture in order to employ an extra output derived from the latent space. Therefore, while the traditional output of the auto-encoder is trained to approximate the input values, the extra output is trained to approximate the next sample of the sequence given to the input of the auto-encoder.

Consider $X$ a sequence such as $X = x_1, x_2, ..., x_n$, the latent space vector $H$ is obtained with the mapping $p_{encoder}(H|X)$ and the traditional output of the auto-encoder is obtained with the mapping $p_{decoder}(\hat{X}|H)$. The extra output added to the auto-encoder model tries to approximate $x_{n+1}$ with the mapping $p_{predictor}(x_{n+1}|H)$.

\begin{figure}[ht]
    \centering
    \includegraphics[scale=0.15]{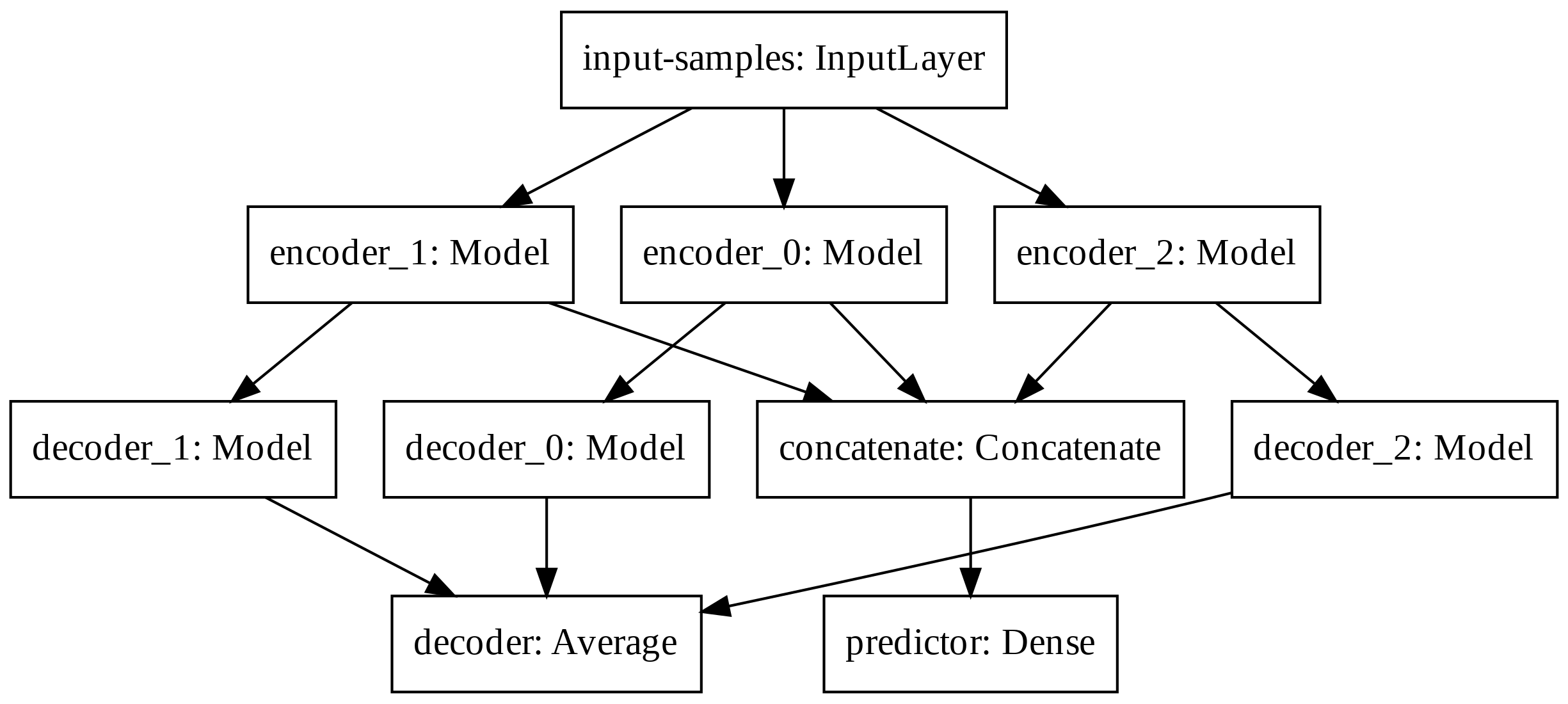}
    \caption{Modified Auto-Encoder architecture}
    \label{fig:mae}
\end{figure}

In order to increase the latent space dimension without increasing the input sequence, we apply 3 auto-encoders in parallel and aggregate their latent space before computing the predictor output. Such Modified Auto-Encoder (MAE) architecture is depicted in Figure \ref{fig:mae}.

The predictor output, the input-samples and the decoder output have 1, 8 and 8 units, respectively. Each encoder, latent space and decoder has 32, 4, and 32 units, respectively. The output of each decoder is averaged to create the total decoder output. The latent spaces of each auto-encoder are concatenated prior the final computation of the predictor output. We train the modified architecture with the mean squared error loss function and the Adam optimizer.

\subsubsection{Data Processing}

Lets consider the epidemic curve a time-series that models the advance of an epidemic by measuring the number of new confirmed cases of Covid-19 on a daily basis. Hence, we first apply a moving average filter of size 3 to deal with the variability of data related to the amount of tests available and delays in reporting between other problems.

We compute the input examples by dividing the whole epidemic curve in overlapped segments of $8$ days, shifted one day from each other, with the $9th$ day being the value to be forecast by the MAE. Each example is normalized by dividing its values by its maximum value. A set of 10 examples is taken from the most advanced places in each cluster to form a batch of examples. 

In order to evaluate the most advanced places in the epidemic timeline, we compute the difference between the number of cases sampled at the day of peak occurrences and the last number of cases reported. If this difference is positive, the number of daily cases started to decrease, meaning that such place passed the peak number of cases and is more advanced in the epidemic timeline.

\subsubsection{Forecasting New Daily Cases}

In order to forecast the Brazilian epidemic curve, we start by applying the same moving average filter of size $7$ to the epidemic curve of the Brazilian states as depicted in details in Section \ref{sec:methods_clustering}, then we perform the forecasting on these clustered time series data in two phases. 

The first phase uses existing data to feed the network, and the forecast value is one-step ahead of the current example. In the second phase, referred to as multi-step ahead, we use the predicted value of the $i-th$ step to forecast the value of the $(i+1)-th$ step. In this way, it is not necessary to have existing data for the second phase of forecasting, allowing us to forecast the epidemic behaviour several days ahead and identify the probable date of the peak number of daily cases, which might indicate a drop in the number of occurrences. Notice that this peak or the end of the pandemic might be subject to some displacements due to problems in the data, so a final step needs to be applied in order to verify the peaks for all states. This is done by fitting a distribution curve on the output data as described next.

\subsection{Final approximation for the Covid-19 curves}

Despite we discuss below the impossibility of finding a curve that mathematically represents the Covid-19 dissemination, the main and most important reason for trying to approximate this curve is that it allows to define useful information such as verifying the peak, and estimating the end of the pandemics. Moreover, it can generate more realistic number of cases to some degree of precision, thus being of importance. To determine the end of the pandemic or the peak are two of these advantages, as it is supposed that epidemics obey certain statistical rules \cite{Baerwolff2020}, to some degree of precision. In this work we verify the peaks after approximating the final predicted curve using some statistical procedure.

In relation to modeling Covid-19 using statistical distributions, it has been discussed that this is a somewhat difficult task. Actually, the Covid-19 curve can not be considered a Gaussian probability distribution \cite{Samos2020}. In fact, it is argued that the shape of a normal distribution is a histogram that is a transformation of probability density against values of a single variable while the Covid-19 contagion curve is a transformation of the values of one variable (confirmed cases) according to a second variable (time). So the curve is not a distributions in the sense usually meant in probability and statistics. Nonetheless, one can visually notice that the curve of daily confirmed cases $\times$ time (day) looks like a distorted Gaussian, and can actually be approximated by some distributions such as the normal (rarely), pearson, logistic, logNormal, and gamma, among others. For the sake of confirming or ratifying the estimated peak, we thus conduct a statistical procedure to the time series data output by the MAE models.

\section{Results}

In the following we present the experimental results for validating the methods introduced in this work. We start by describing some results found in the literature for traditional approaches followed by the LSTM results, as a comparison discussion over these approaches will be further conducted. Then results of the clustering procedure are shown in order to validate and clear the approach used. Finally, we present the results obtained with the Modified Auto-Encoder model to forecast the epidemic curve of Covid-19 in Brazil, as well as the approximated distribution curve confirming the peaks obtained by the above fitting procedure, for all Brazilian states. The numbers for Brazil, which are of straight interest to the population, are presented for some of the states and are available at \url{www.natalnet.br/covid}. We notice one more time that these numbers are predictions, as so they might get different from the real numbers as the pandemic dynamics evolves.

\subsection{SIR, SEIR and SIAD results}

Results with SIR and SEIR can be found including several applications running on the Internet \cite{Lyra2020Site}. Before entering these results, we note that we could not find accurate results on Covid-19 long-term dynamics prediction for these methods, up to date. However any approximation is useful at this time, as long-term forecasts may help managers to discuss different types of confinement policies \cite{bastos2020modeling}, and it can help to come up with an estimate of the optimal date to end the confinement policy. For example, the preliminary results reported by Bastos and Cajueiro in April 1st \cite{Bastos2020A}, using SIR and SIAS, are far from being accurate. Indeed, they suggest that 30 million people will get infected in Brazil on May 11th (pandemic predicted peak) in the less infected people situation. Later, in their revised work evolving to SID and SIASD approaches \cite{bastos2020modeling}, the results are better, however not precise yet.

\begin{figure}
    \centering
    \includegraphics[width=\textwidth]{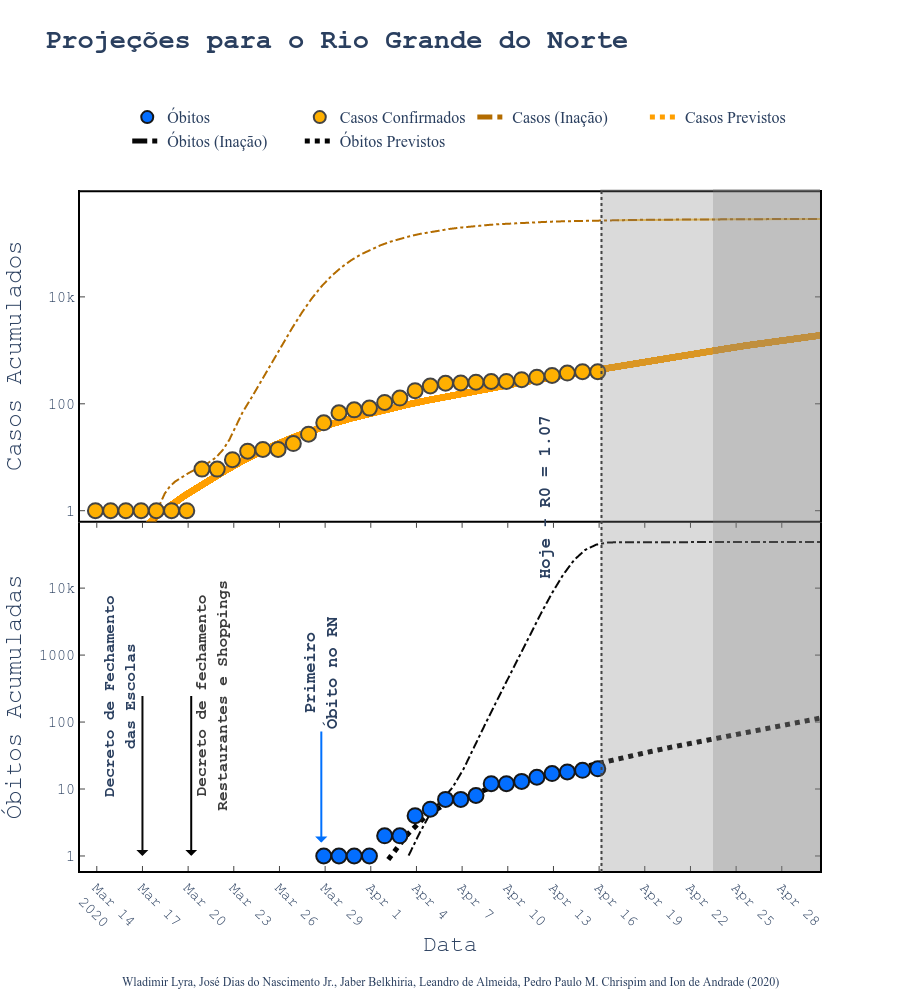}
    \caption{Projections for Rio Grande do Norte state (at the northeast of Brazil) \cite{Lyra2020}. Figure printed out from the web application running at http://astro.dfte.ufrn.br/html/Cliente/COVID19.php. Acessed on May 4th.}
    \label{fig:dfteRN}
\end{figure}

\begin{figure}
    \centering
    \includegraphics[width=\textwidth]{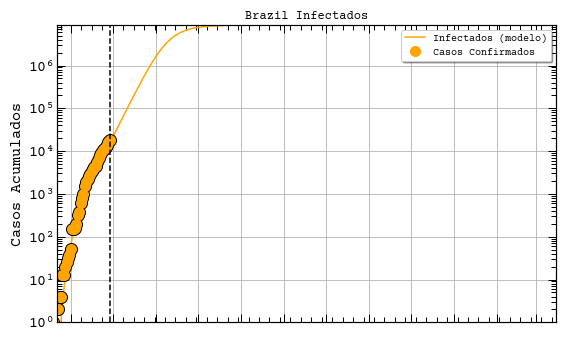}
    \caption{Projections for Brazil with adapted SEIR model \cite{Lyra2020}, extracted from http://astro.dfte.ufrn.br/html/Cliente/COVID19.php. Acessed on May 4th.}
    \label{fig:dfteBR2}
\end{figure}

Referring to the Northeast (our main interest region), a web site for monitoring of COVID of UFRN \cite{Lyra2020Site} is used by the Govern, which is based on a modified SEIR accounting for social distancing rules \cite{Lyra2020}. According to them, the epidemic started on March 1st and the symptomatic cases are predicted to end on July 1st. The peak of symptomatic people is predicted for May 17th with 20 million. In a more detailed look of the web page (at May 4th), a php application says that Brazilian state RN could have 2,039 confirmed cases on April 30th 2020, following the current scenario of social distancing, as shown in Figure~\ref{fig:dfteRN}, available and printed from the website. Actually, we have had an outcome of 1,297 confirmed cases for that day, not so bad (less than 50 \% error). For Brazil, their prediction was about 742 thousand confirmed cases by the same day. The actual number was 87,187 confirmed cases (according to the Health Ministry of Brazil - MS). This is a greater disparity between values. This picture is shown in Figure~\ref{fig:dfteBR2}, also available from their website (\url{http://astro.dfte.ufrn.br/html/Cliente/COVID19.php}). Again, these predictions, besides a bit exaggerated, are useful for the authorities to take decisions, reporting the worst cases of the pandemic, most of the time. As it will be shown, our results tend to be a little more humble than the ones reported on this site and we sincerely hope that our predictions are still exaggerated.

\subsection{Preliminary results with LSTM}

\begin{figure}
\centering
\begin{subfigure}[b]{\textwidth}
    \includegraphics[width=\textwidth]{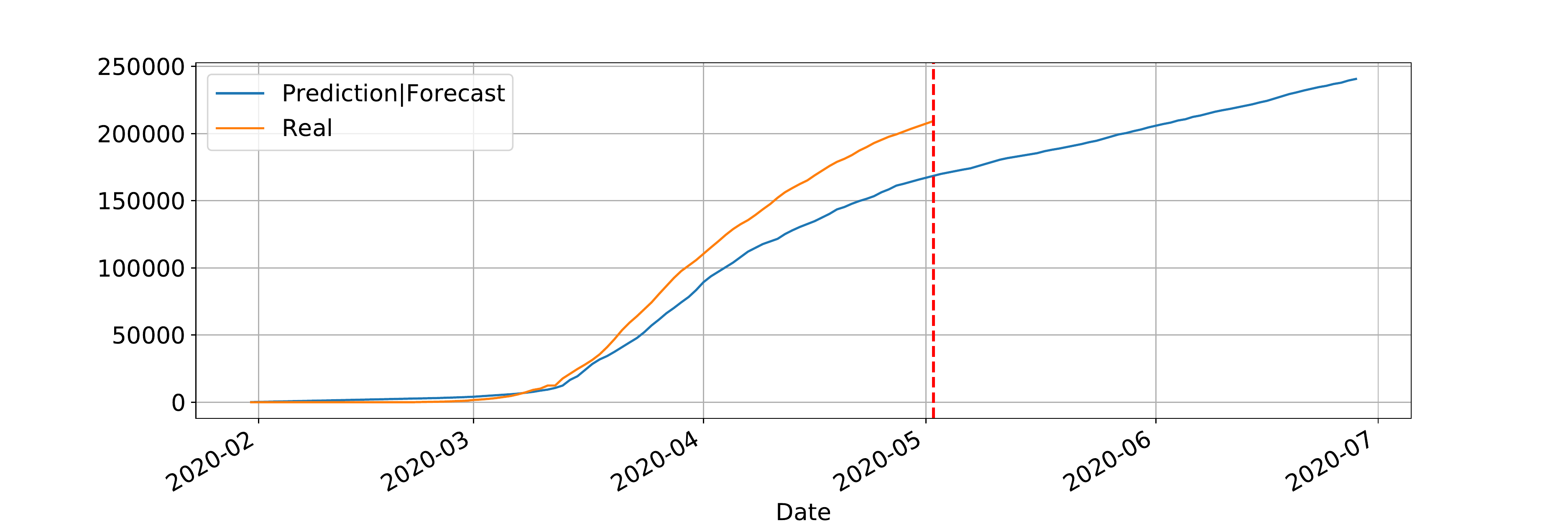}
    \caption{}
    \label{fig:italy_tatal_lstm}
\end{subfigure}
\begin{subfigure}[b]{\textwidth}
    \includegraphics[width=\textwidth]{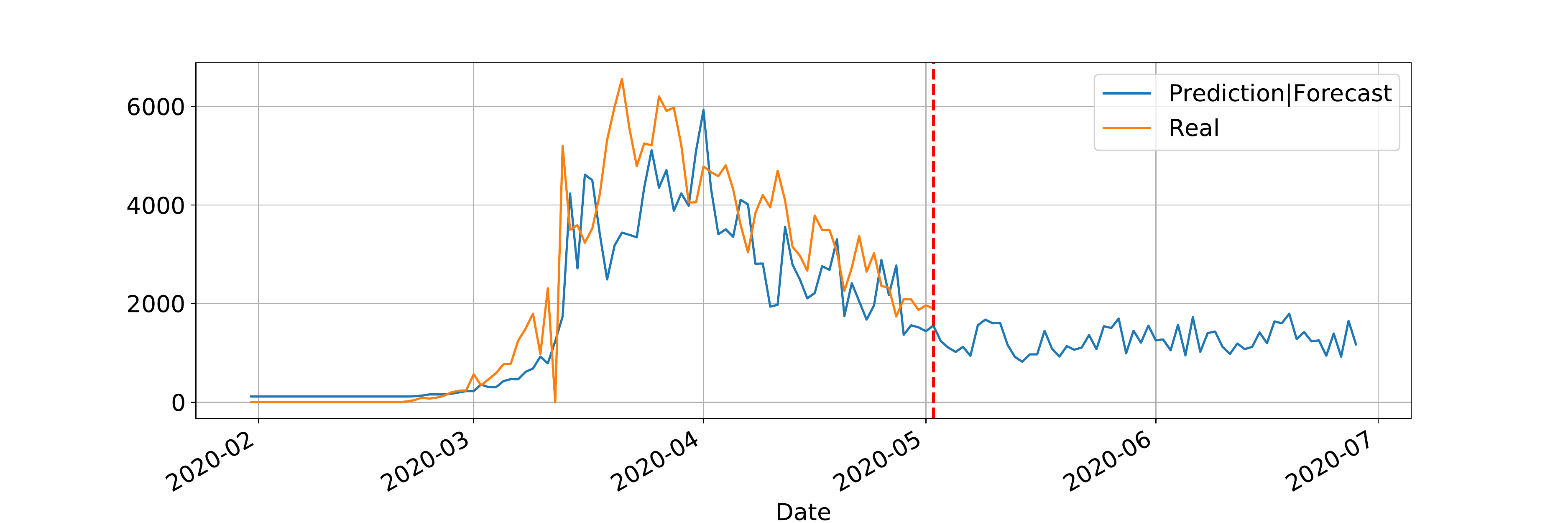}
    \caption{}
    \label{fig:italy_daily_lstm}
\end{subfigure}
\caption{Predictions and forecasting to Italy on Covid-19 cumulative (\ref{fig:italy_tatal_lstm}) and daily (\ref{fig:italy_daily_lstm})}
\label{fig:italy_lstm}
\end{figure}

\begin{figure}
\centering
\begin{subfigure}[b]{\textwidth}
    \includegraphics[width=\textwidth]{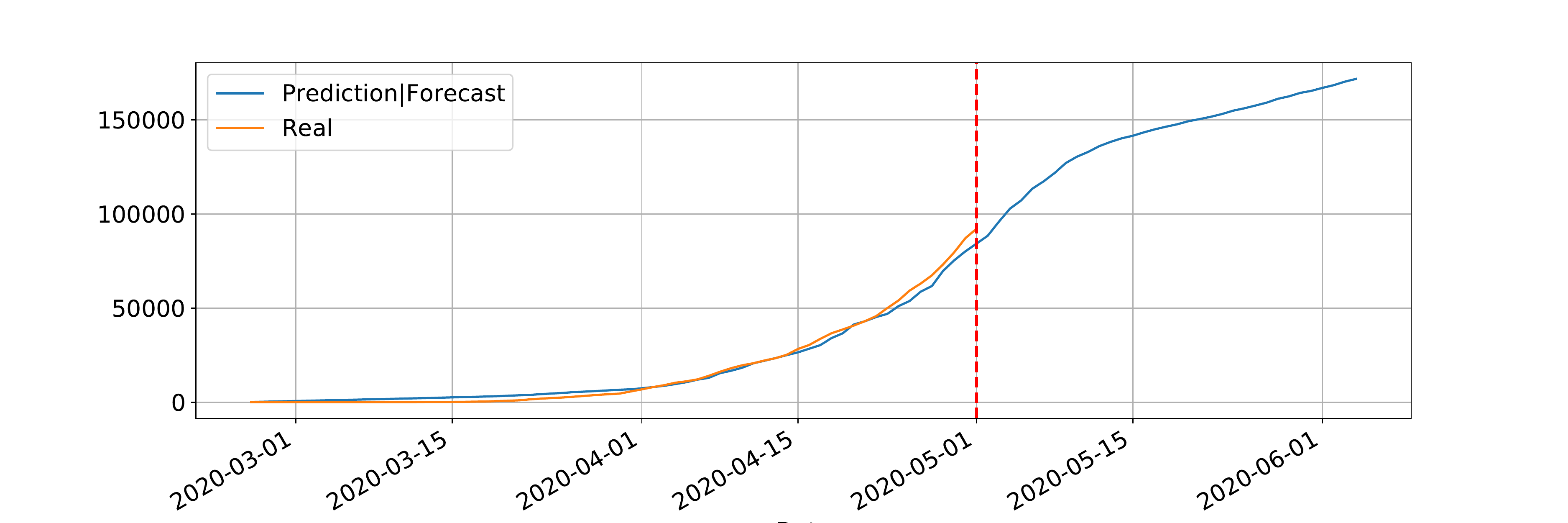}
    \caption{}
    \label{fig:brazil_tatal_lstm}
\end{subfigure}
\begin{subfigure}[b]{\textwidth}
    \includegraphics[width=\textwidth]{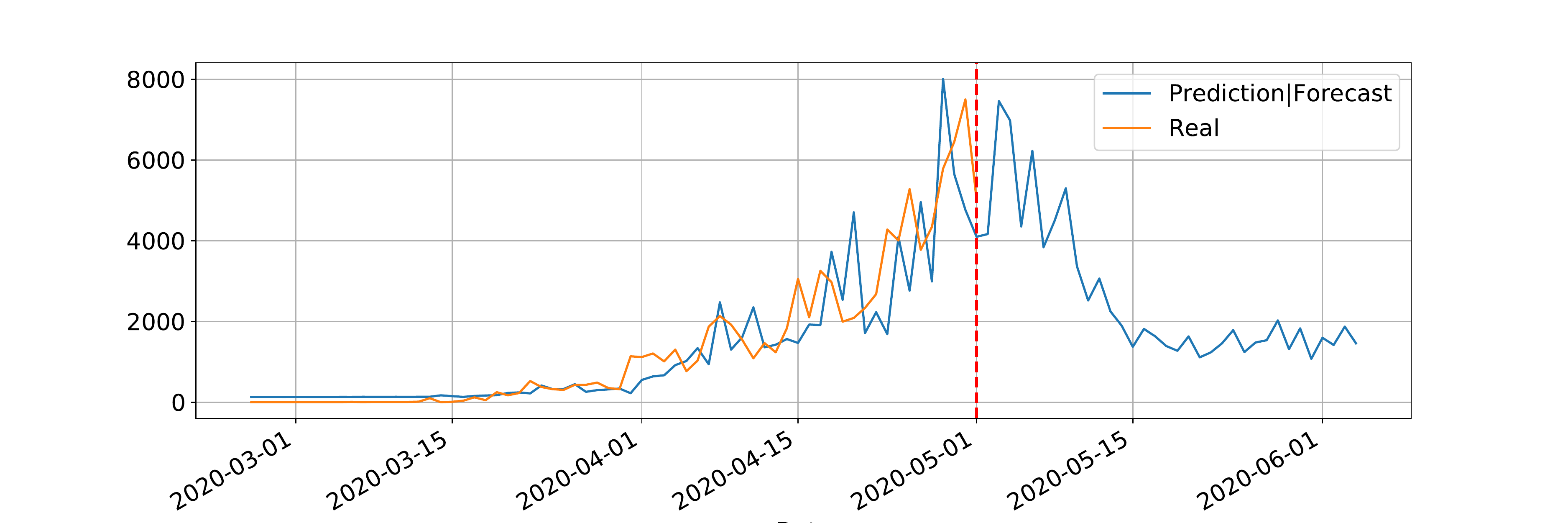}
    \caption{}
    \label{fig:brazil_daily_lstm}
\end{subfigure}
\caption{Predictions and forecasting to Brazil on Covid-19 cumulative (\ref{fig:brazil_tatal_lstm}) and daily (\ref{fig:brazil_daily_lstm})}
\label{fig:brazil_lstm}
\end{figure}

\begin{figure}
\centering
\begin{subfigure}[b]{\textwidth}
    \includegraphics[width=\textwidth]{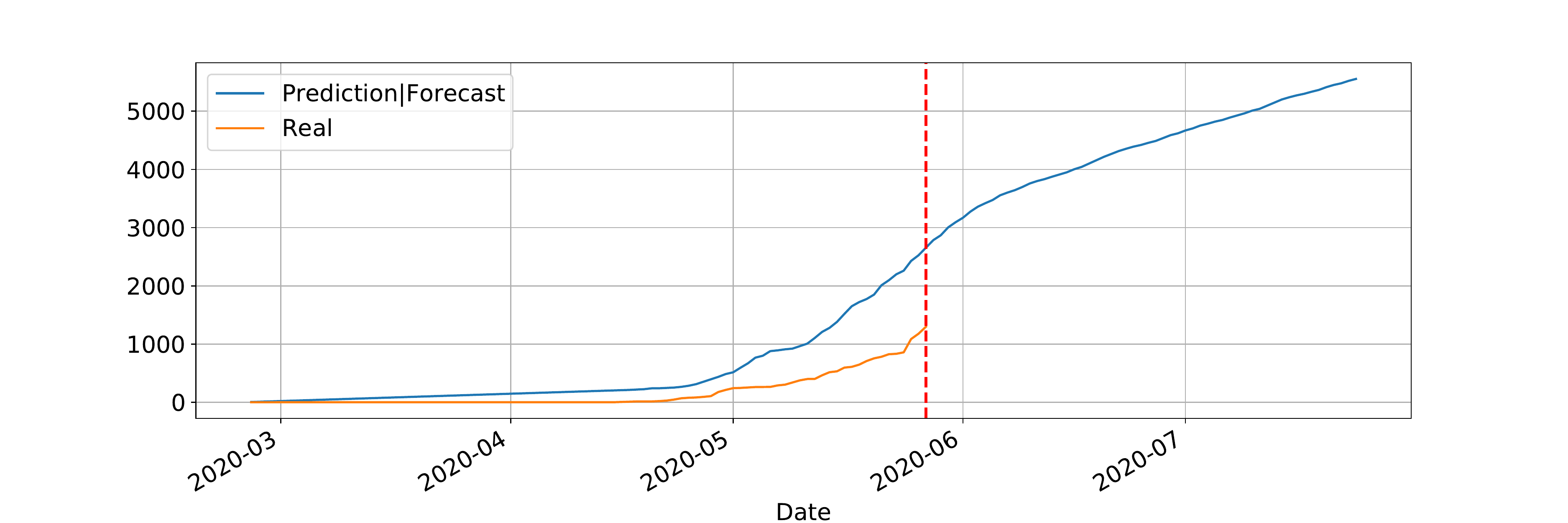}
    \caption{}
    \label{fig:rn_tatal_lstm}
\end{subfigure}
\begin{subfigure}[b]{\textwidth}
    \includegraphics[width=\textwidth]{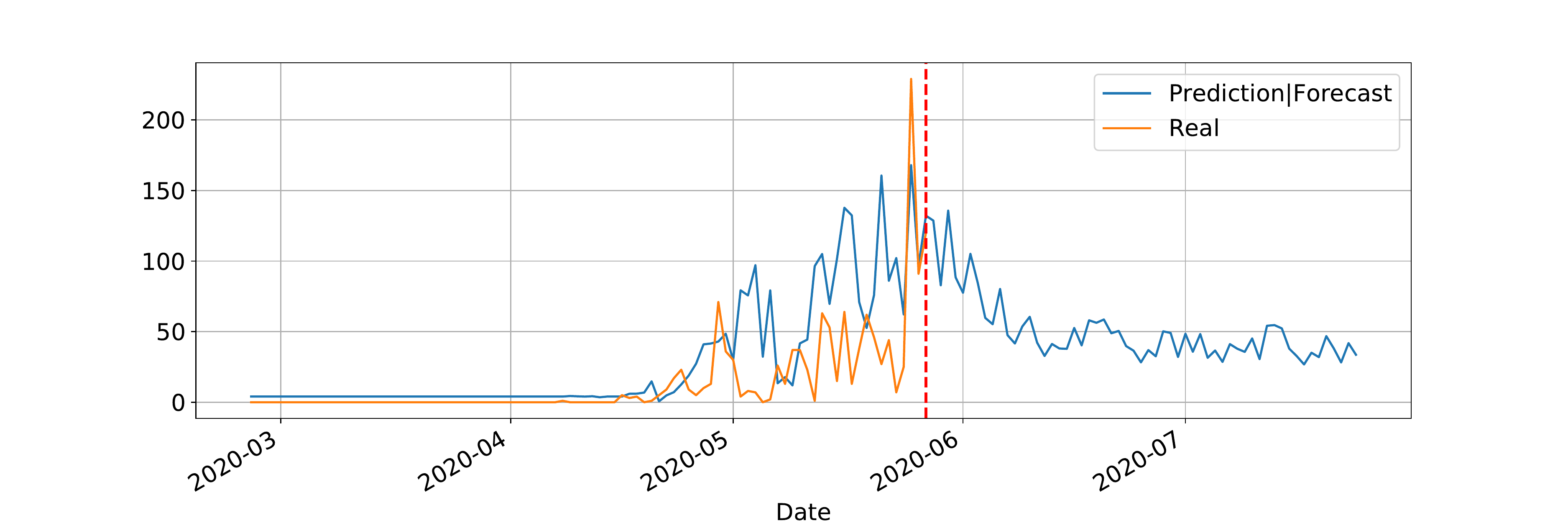}
    \caption{}
    \label{fig:rn_daily_lstm}
\end{subfigure}
\caption{Predictions and forecasting to RN on Covid-19 cumulative (\ref{fig:rn_tatal_lstm}) and daily (\ref{fig:rn_daily_lstm})}
\label{fig:rn_lstm}
\end{figure}

In our initial studies towards data driven approaches, we tested the possibility of using the LSTM-type RNN for determining Covid-19 dynamics for our region of most interest (Brazilian state RN), but it did not work as expected due to several factors. Mainly the under-notification of data made available by the governments of Brazil and its Federate States. Hence, more work is necessary for improving and testing this model, and adjusting it to predict the dynamics of the pandemic, including its various parameters. The problems with this architecture applied to COVID-19 data will be further discussed in the Section~\ref{sec:discussion}. Basically, the main drawback of the model is its inability to reset at certain time and bringing the values to zero (or close). Anyway, we used the LSTM-SAE to forecast three different places, with different phases of the disease: Italy (Figure~\ref{fig:italy_lstm}), which contamination curve is starting to decrease; Brazil (Figure~\ref{fig:brazil_lstm}), which is approaching the peak; and the state of Rio Grande do Norte (Figure~\ref{fig:rn_lstm}) that is about to reach the peak.

{}

{}

{}

Although not responding perfectly, we notice some LSTM important features that can be seen on the charts. One is that the LSTM-SAE model could stabilize over time. By analyzing the daily results, the other LSTM models cannot return to zero and keep oscillating around some positive value. Because of this, when the value is accumulated it always increases. This issue is more apparent when the model is used for countries or regions that did not stabilize their cases. These limitations can be associated to the non-linearity of data, among other issues. Another point is that, as presented in previous work \cite{Sagheer2020}, the LSTM-SAE addresses the input data randomization in the LSTM block. The encoder-decoder model trained first feeds the hidden layer with initialization weights. It is possible that because of that, the LSTM-SAE architecture presents the best results. More complete results using LSTM can be found at \url{www.natalnet.br/covid}.

\subsection{Checking the clustering results (input to MAE)}
\label{sec:results_clustering}

\begin{figure}
    \centering
    \includegraphics[trim={100 80 80 0},clip,width=0.9\textwidth]{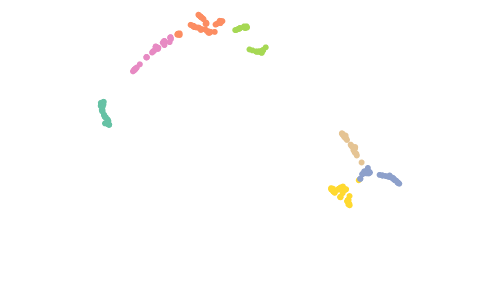}
    \caption{2D UMAP embedding of the different countries and states studied. The colors represents different clusters generated using Affinity Propagation.}
    \label{fig:umap}
\end{figure}

Before showing MAE results, this subsection presents and discusses the results from the preliminary clustering necessary for the better performance of MAE, which was presented in Section~\ref{sec:methods_clustering}. To evaluate qualitatively the clusters obtained, lets use the 2D UMAP embedding shown in Figure~\ref{fig:umap}. Seven clusters were formed, and overall, they seem rather compact and distinct from each others. Although there is a slight overlapping between some pairs of clusters, this plot suggests that there was actually well defined groups within the different countries/regions, probably reflecting the types of actions taken by the governments to react to the early signs of the pandemic. Notice that we suppose and believe that countries from the same clusters should follow similar contamination curves.

{}

\begin{figure}
	\centering
	\begin{subfigure}{0.5\textwidth}
	\centering
	\includegraphics[width=\textwidth]{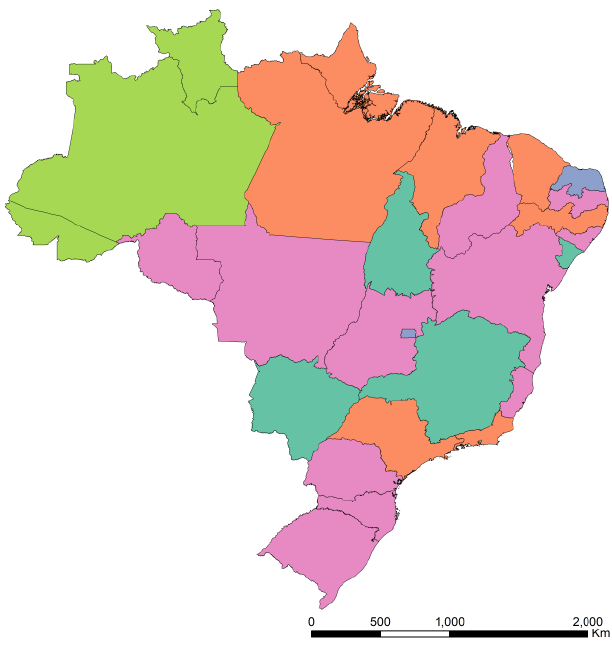}
    \caption{Brazilian states}
	\end{subfigure}
	
	\begin{subfigure}{\textwidth}
	\centering
	\includegraphics[width=\textwidth]{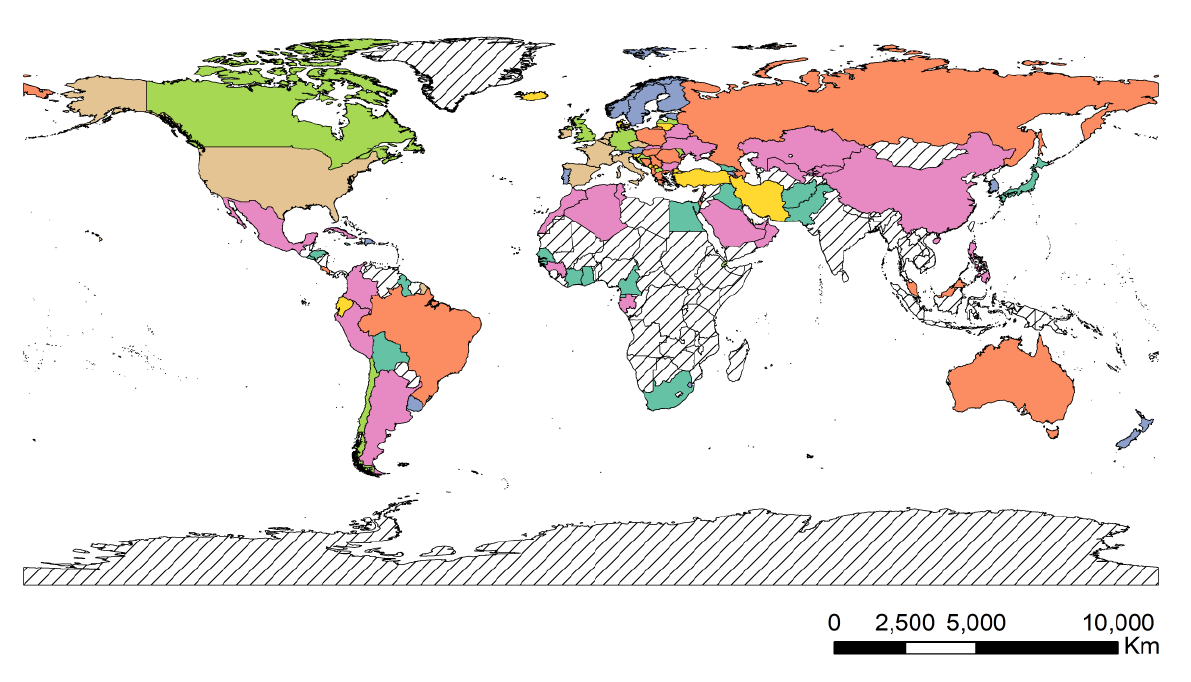}
	\caption{World} 
	\end{subfigure}

	\caption{Clusters assignment of the different Brazilian states and world countries.}
    \label{fig:cluster_maps}
\end{figure}

In order to visualize this preliminary classification and get some insight for Brazilian states, a map of Brazil representing the clusters is shown at Figure~\ref{fig:cluster_maps}. In that map we separate in the same color the states and countries that presented a similar reaction to the outbreak of Covid-19. The countries which are represented by hatches in the maps were either not sufficiently advanced at the time of the study or their time series produced numerical instabilities during feature computation. The states from USA, Australia, Italia, China and Canada appear in the different clusters but are not represented in the maps to improve readability of the paper. The full results of the cluster assignment used in the training process can be found at \url{www.natalnet.br/covid}.

{}

\begin{figure}
    \centering
    \includegraphics[trim={60 0 60 0},clip,width=\textwidth]{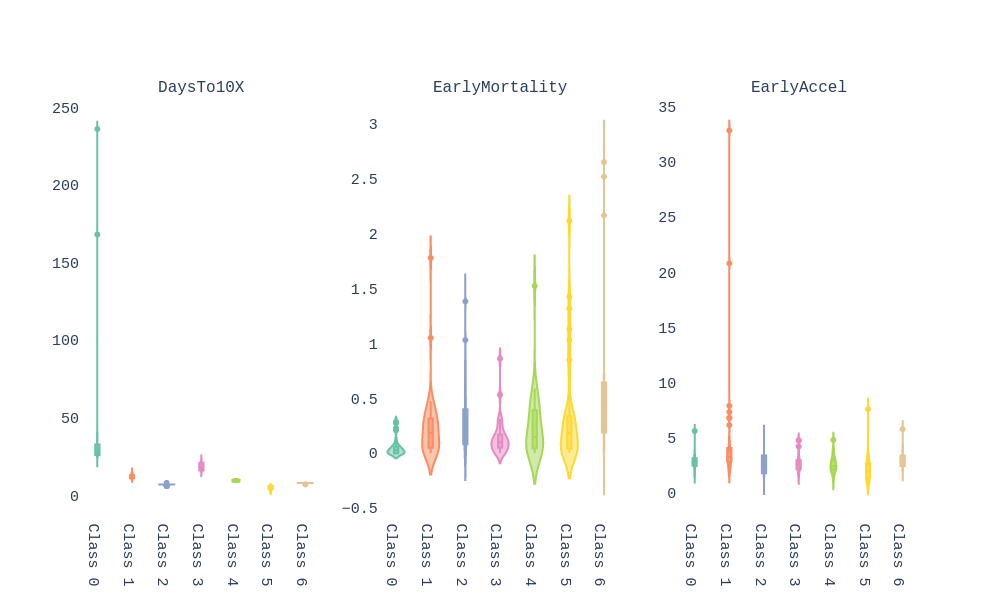}
    \caption{Violin plots representing the values taken by the different features for each groups obtained after UMAP + Affinity Propagation clustering.}
    \label{fig:violin}
\end{figure}

Finally, the values of the features of the different groups are presented in the form of violin plot on Figure~\ref{fig:violin}. We can see, for example that cluster 0 gathers the countries with higher \emph{Days until 10x}, meaning that these countries/regions managed to contain the contagion early. In turn, Brazil belongs to cluster 1, which contains countries with high early acceleration and above average mortality rate.

{}

We conclude this section by underlining the fact that the colors representing the clusters used for Figures~\ref{fig:umap}, \ref{fig:cluster_maps} and \ref{fig:violin} are matching, meaning that a country in yellow on the map belongs to the yellow cluster on the UMAP plot and its statistics can be seen in yellow on the violin plot. In addition, we believe that all major centers of Covid-19 are represented in these maps, which provides sufficient material to train our models.

\subsection{MAE Results}

This Section presents the results obtained by applying the MAE architecture model to forecast the Covid-19 epidemic curves of Brazilian states of each cluster. Therefore, for each cluster, a MAE model was trained with the $10$ most advanced countries of the cluster with the data available up to the day of this study, and the epidemic curves for the Brazilian states of the cluster were forecast. We note, however, that the Brazilian states were only on clusters $0$, $1$, $2$ and $3$.

Here, we depict one state for each cluster. For an interactive visualization of all Brazilian states you may refer to \url{https://www.natalnet.br/covid}.

\begin{figure}[h!]
\begin{subfigure}{\textwidth}
    \centering
    \includegraphics[width=\textwidth]{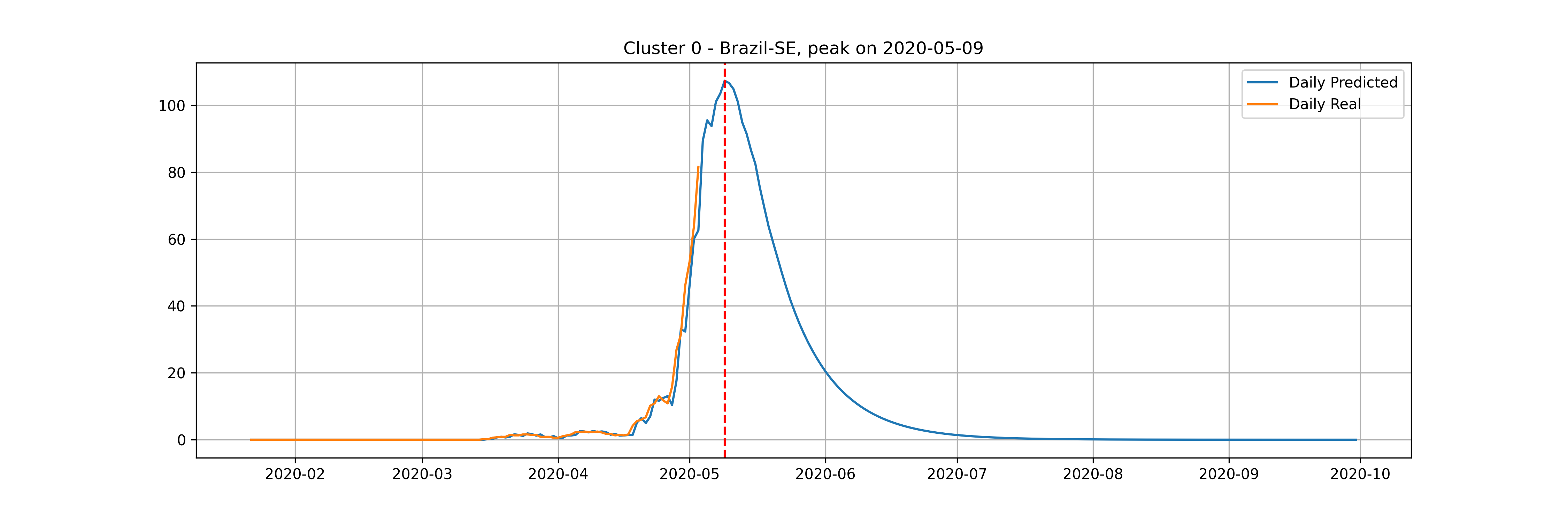}
    \caption{Daily cases for Sergipe State.}
\end{subfigure}
\newline
\begin{subfigure}{\textwidth}
    \centering
    \includegraphics[width=\textwidth]{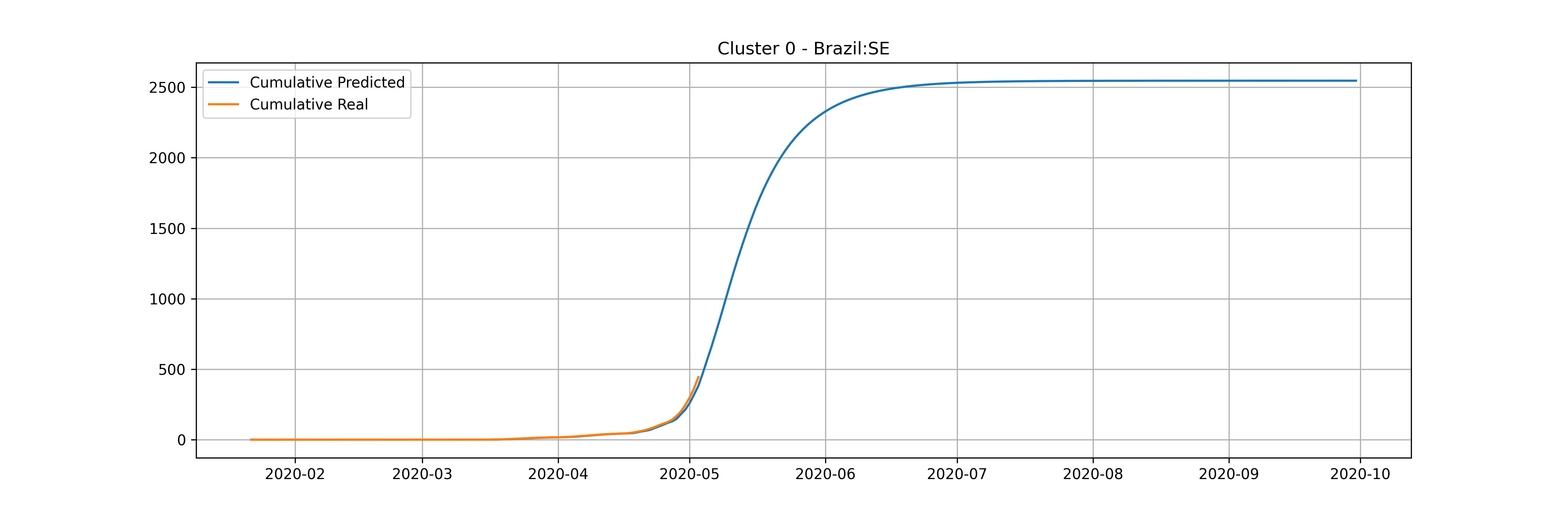}
    \caption{Cumulative cases for Sergipe State.}
\end{subfigure}
\caption{Daily and Cumulative cases for Sergipe State from the Cluster $0$.}
\label{fig:c0}
\end{figure}

In Figure \ref{fig:c0}, the daily and cumulative epidemic curves for the Sergipe state is displayed. The peak for the Sergipe state is predicted to happen on May $9$ and should reach up to $2546$ total number of cases at the mid of July.

\begin{figure}[h!]
\begin{subfigure}{\textwidth}
    \centering
    \includegraphics[width=\textwidth]{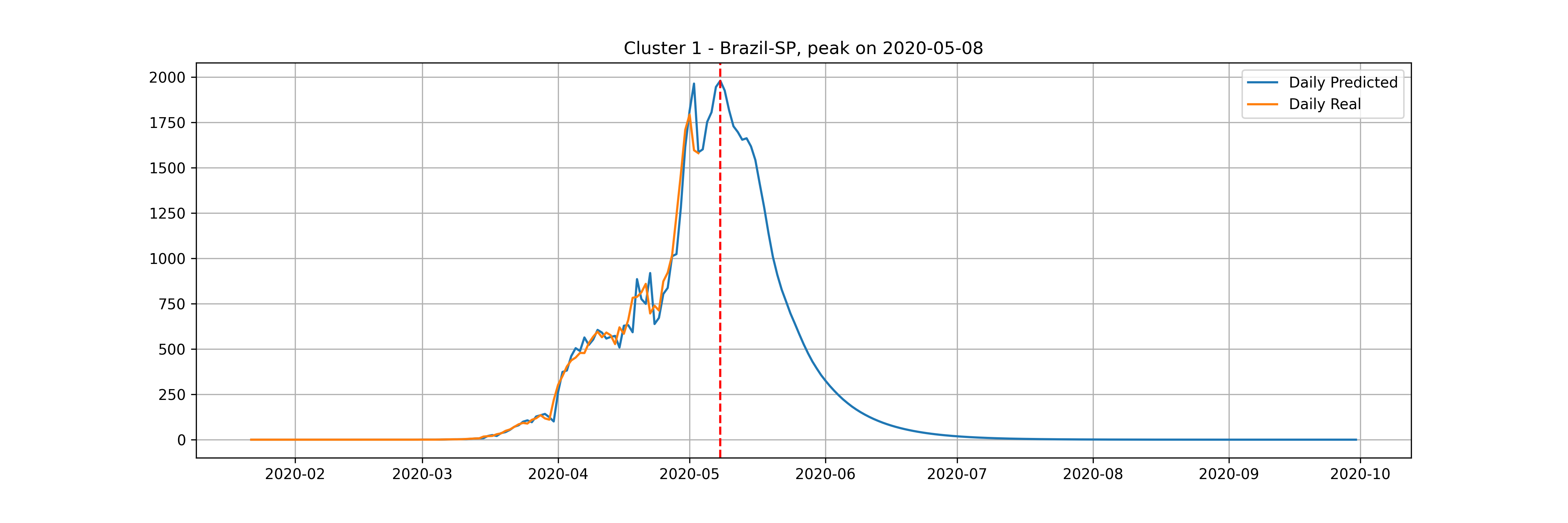}
    \caption{Daily cases for São Paulo State.}
\end{subfigure}
\newline
\begin{subfigure}{\textwidth}
    \centering
    \includegraphics[width=\textwidth]{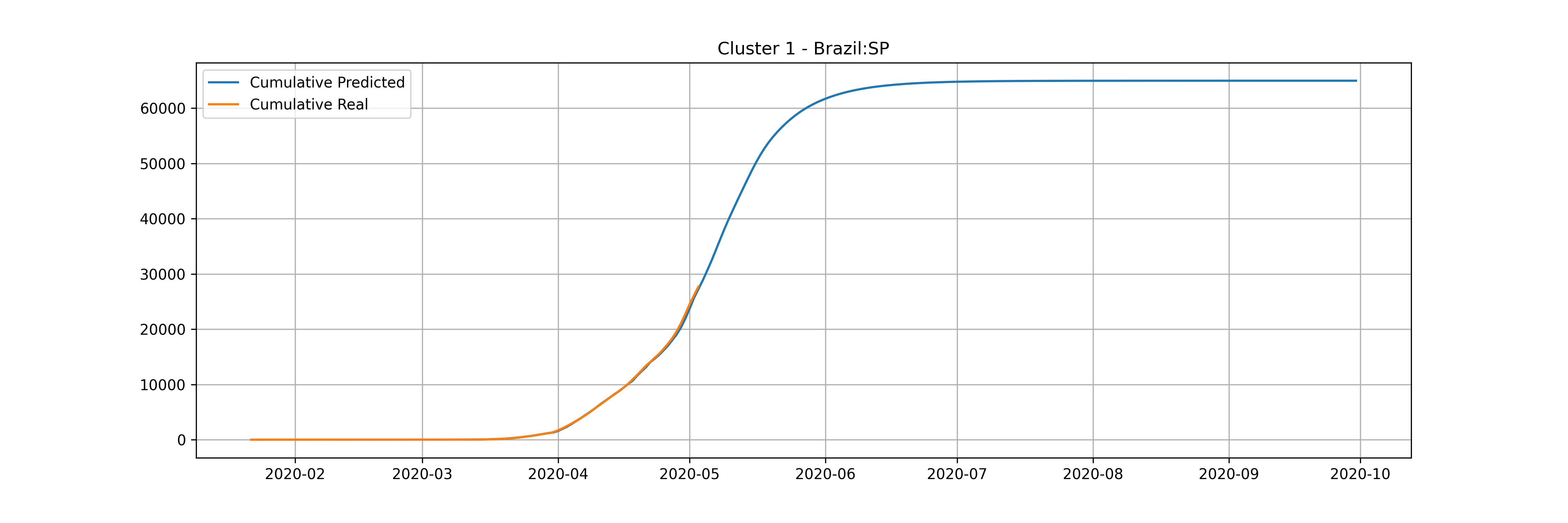}
    \caption{Cumulative cases for São Paulo State.}
\end{subfigure}
\caption{Daily and Cumulative cases for São Paulo State from the Cluster $1$.}
\label{fig:c1}
\end{figure}

Figure \ref{fig:c1} depicts the epidemic curves for the S\~ao Paulo state. In this case, we predict the peak number of cases for May $8$ and that the state would reach a total of $64,984$ cases at mid-July.

\begin{figure}[h!]
\begin{subfigure}{\textwidth}
    \centering
    \includegraphics[width=\textwidth]{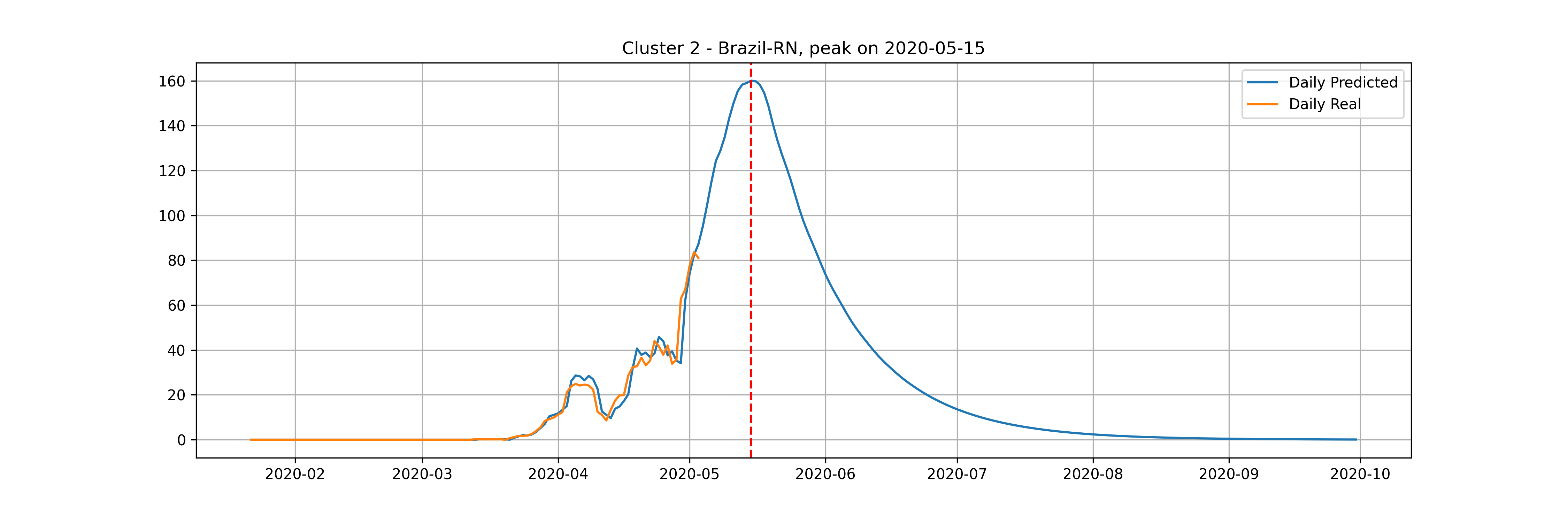}
    \caption{Daily cases for Rio Grande do Norte State.}
\end{subfigure}
\newline
\begin{subfigure}{\textwidth}
    \centering
    \includegraphics[width=\textwidth]{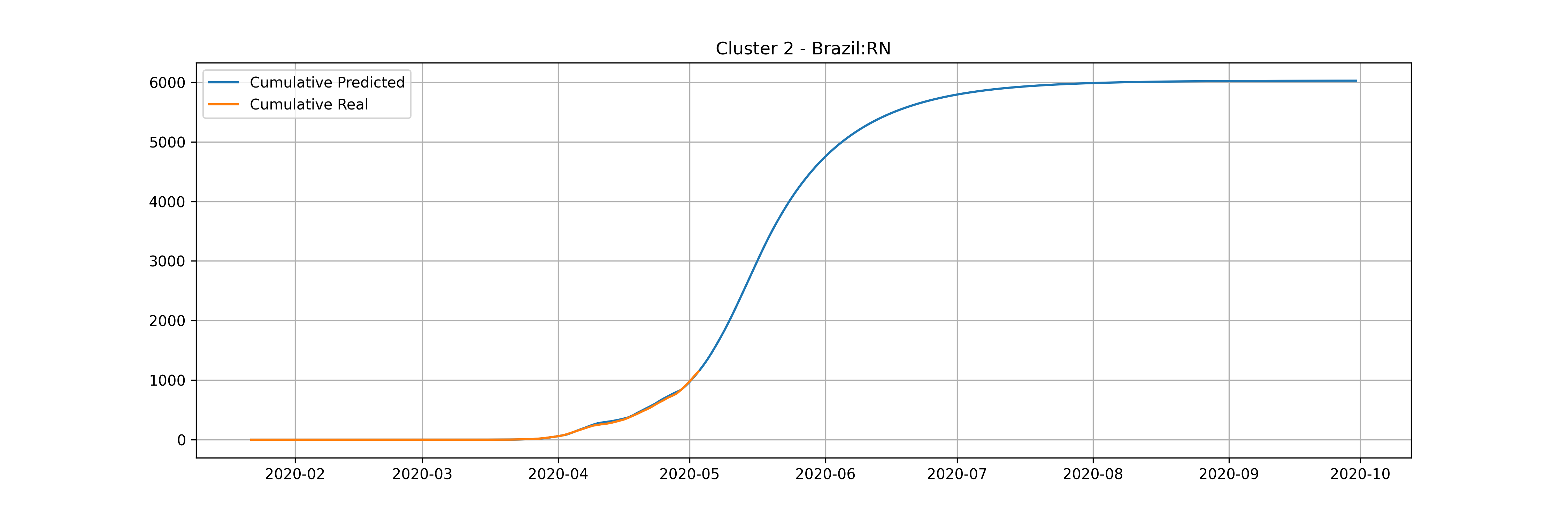}
    \caption{Cumulative cases for Rio Grande do Norte State.}
\end{subfigure}
\caption{Daily and Cumulative cases for Rio Grande do Norte State from the Cluster $2$.}
\label{fig:c2}
\end{figure}

The epidemic curves for the Rio Grande do Norte state is depicted in Figure~\ref{fig:c2}. The peak occurrence in daily cases is predicted to happen in May $15$ and should reach up to $6025$ at the end of August.

For the last cluster, we depict the epidemic curves of the state of Santa Catarina. The peak occurrence is predicted to happen in May $16$ and should reach $15329$ cases at the end of September.

\begin{figure}[h!]
\begin{subfigure}{\textwidth}
    \centering
    \includegraphics[width=\textwidth]{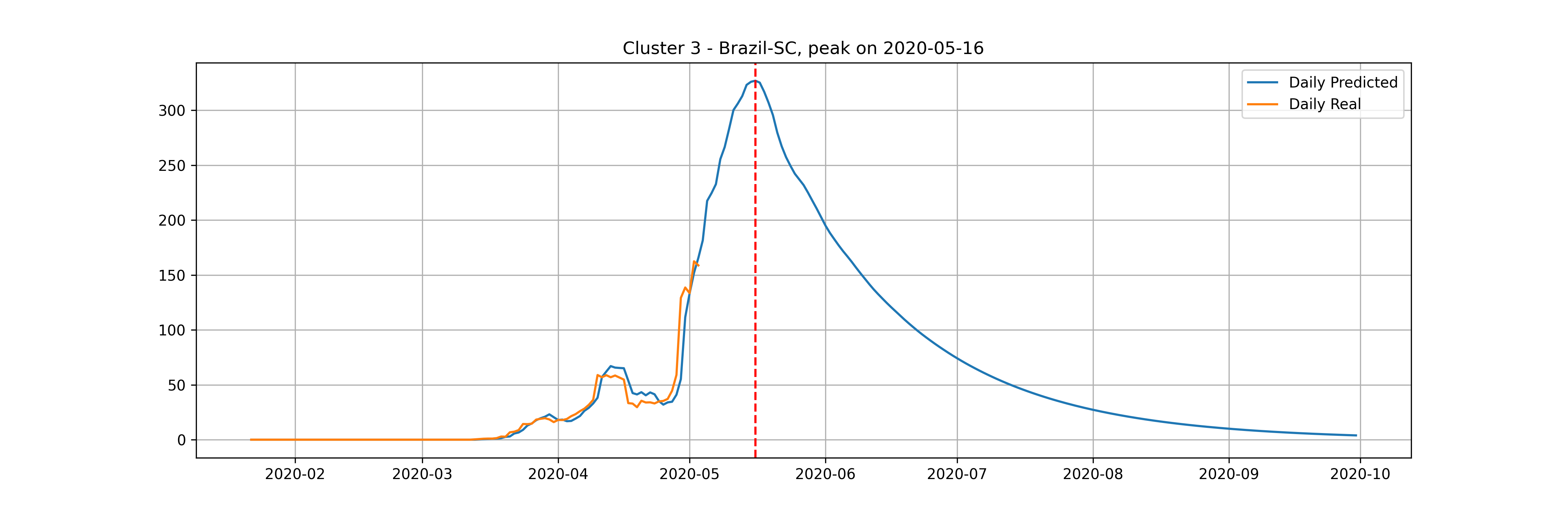}
    \caption{Daily cases for Santa Catarina State.}
\end{subfigure}
\newline
\begin{subfigure}{\textwidth}
    \centering
    \includegraphics[width=\textwidth]{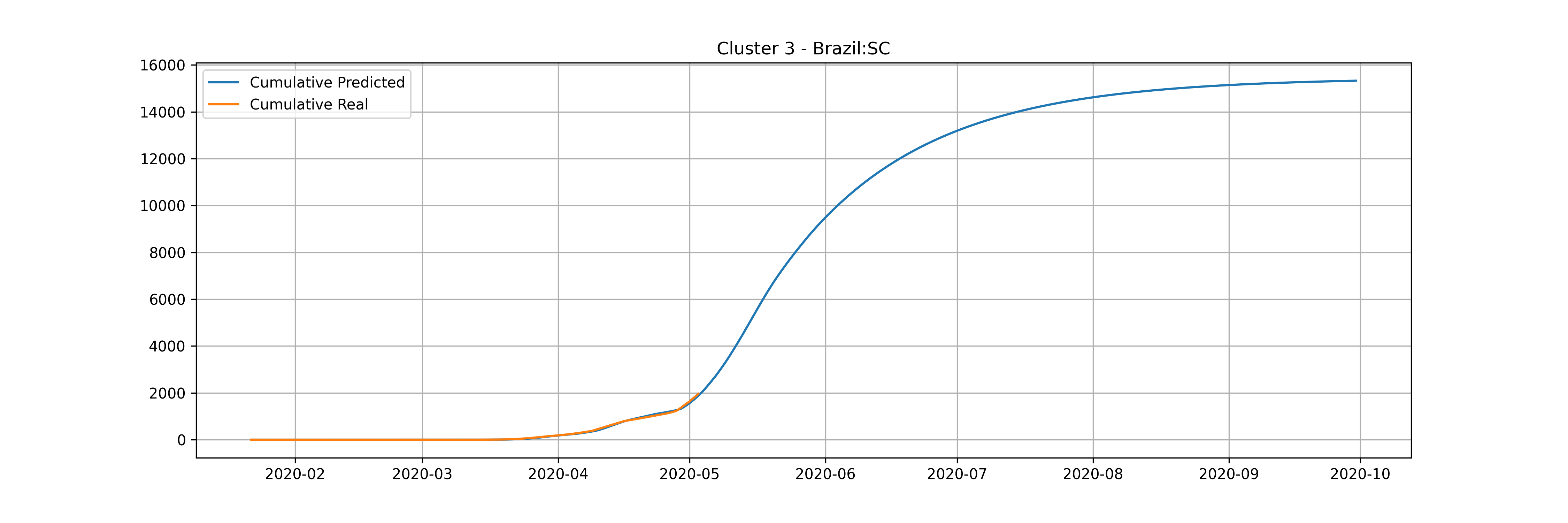}
    \caption{Cumulative cases for Santa Catarina State.}
\end{subfigure}
\caption{Daily and Cumulative cases for Santa Catarina State from the Cluster $3$.}
\label{fig:c3}
\end{figure}

From the epidemic curves illustrated above, we verify that each state has its behavior associated to the cluster it belongs. States from cluster $0$ generally presents a steep peak but a very low number of daily cases, indicating that the epidemic is starting and evolving fast but will not present an elevated number of daily cases.

The cluster $1$ presents a different behaviour. Generally, states from cluster $1$ presents a steep peak with an elevated number of daily cases, meaning that the transmission dynamics is happening much faster than cluster $0$. In the meantime, the predictions show that states from cluster $1$ are close to reach the peak number of occurrence of daily cases and should have its occurrence of daily cases decaying very fast.

The states from cluster $2$ present a slower rate of transmission dynamics if compared to states from the cluster $1$, and according to the date expected for the peak number of daily cases, these states still did not reach the peak number of occurrences.

States from cluster $3$ present the slowest transmission dynamics and tends to have their number of daily cases decaying slowly. 

\begin{figure}
    \centering
    \includegraphics[width=\textwidth]{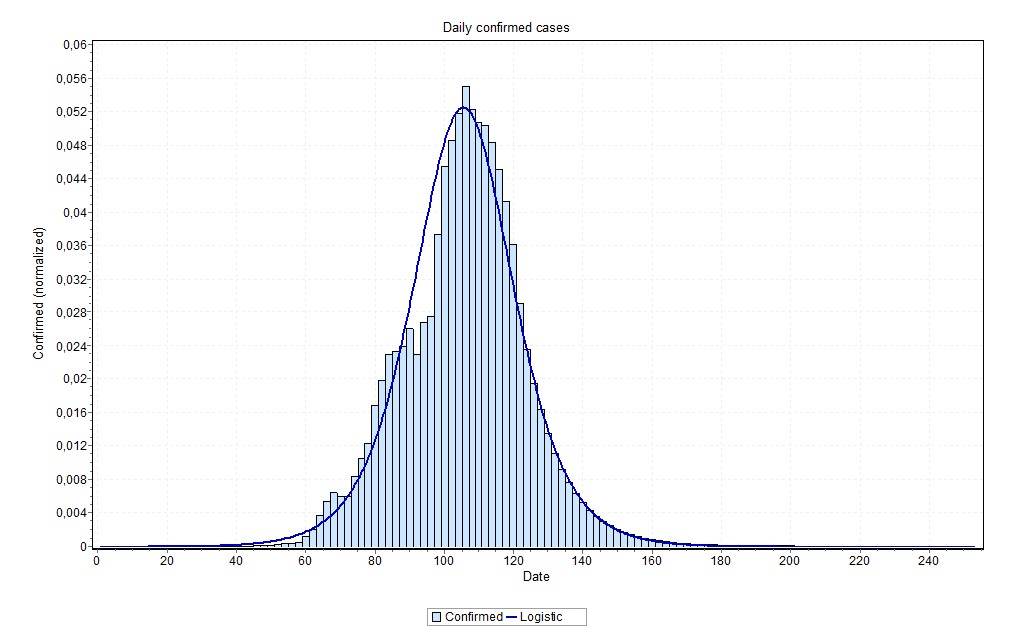}
    \caption{Curve fitting for Rio de Janeiro state (logNormal model was the best fit) with peak is indicated on May 5th, 2020.}
    \label{fig:fittingRJ}
\end{figure}

\begin{figure}
    \centering
    \includegraphics[width=\textwidth]{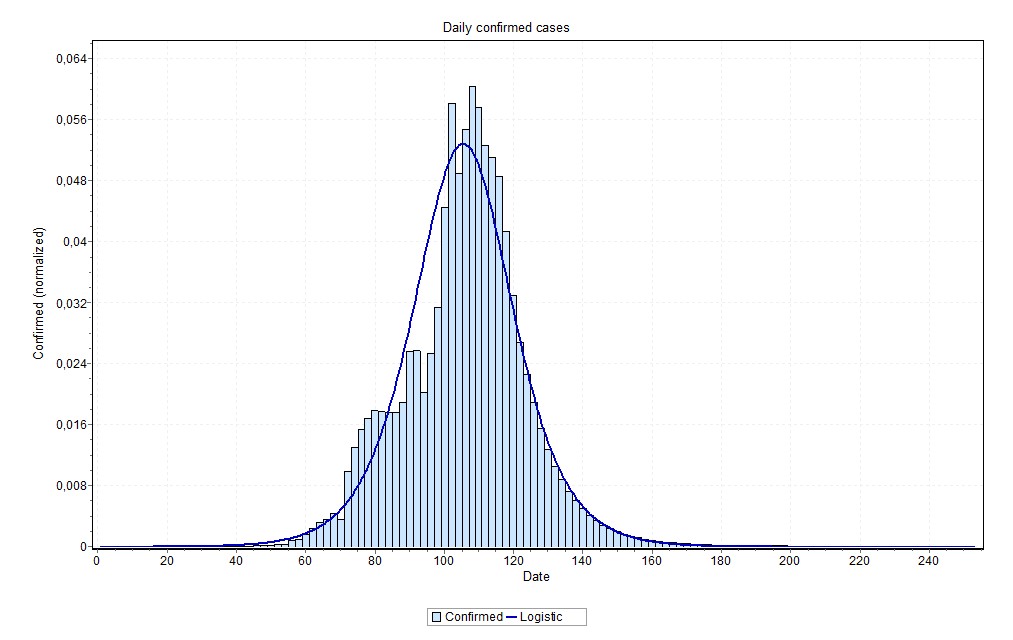}
    \caption{Curve fitting for S\~ao Paulo state (logistic model was the best fit) with peak is indicated on May 6th, 2020.}
    \label{fig:fittingSP}
\end{figure}

We also indicate, in Table~\ref{tab:tb1}, the date of the peak occurrence of cases, the date that it will reach $97\%$ of the total number of cases, the total number of cases and a peak occurrence date obtained by fitting a probability distribution to the predicted curves as well as the curve used in the probability distribution fitting process. Examples of these curves are shown in Figures \ref{fig:fittingRJ} and \ref{fig:fittingSP}, for the states of Rio de Janeiro and S\~ao Paulo, respectively, ratifying the peaks shown in Table \ref{tab:tb1}. The other states curves can be found at \url{www.natalnet.br/covid}.

\begin{table}[h!]\centering
\caption{Peak occurrences for each state predicted by the MAE Model and by a distribution probability. We also indicate the total number of cases expected by the MAE prediction and the day that it'll reach $97\%$ of the total number of cases.}
\scriptsize
\begin{tabular}{lrrrrrr}\toprule
\textbf{State} &\textbf{Predicted by MAE} &\textbf{Curve fit peak} &\textbf{Best curve} &\textbf{Total} &\textbf{97\% of Total} \\\midrule
TO &2020-05-10 &2020-05-10 &Pearson &846 &2020-06-13 \\
SE &2020-05-09 &2020-05-10 &Pearson &2546 &2020-06-13 \\
MG &2020-05-04 &2020-04-30 &Logistic &2992 &2020-06-03 \\
MS &2020-04-25 &2020-04-24 &Pearson &327 &2020-05-28 \\
PA &2020-05-09 &2020-05-10 &Lognormal &10332 &2020-06-11 \\
AP &2020-05-12 &2020-05-12 &Logistic &5172 &2020-06-15 \\
MA &2020-05-07 &2020-05-07 &Lognormal &9684 &2020-06-10 \\
CE &2020-04-30 &2020-04-28 &Pearson &11556 &2020-05-29 \\
PE &2020-05-04 &2020-05-05 &Lognormal &18210 &2020-06-08 \\
RJ &2020-05-05 &2020-05-05 &Lognormal &21587 &2020-06-07 \\
SP &2020-05-08 &2020-05-06 &Logistic &64984 &2020-06-07 \\
RN &2020-05-15 &2020-05-13 &Lognormal &6025 &2020-07-06 \\
DF &2020-05-16 &2020-05-17 &Logistic &6347 &2020-07-06 \\
RO &2020-05-12 &2020-05-14 &Pearson &3061 &2020-08-10 \\
PI &2020-05-16 &2020-05-19 &Pearson &4974 &2020-08-13 \\
PB &2020-05-16 &2020-05-21 &Pearson &8765 &2020-08-14 \\
AL &2020-05-11 &2020-05-19 &Pearson &8119 &2020-08-11 \\
BA &2020-05-07 &2020-05-08 &Pearson &8945 &2020-08-04 \\
ES &2020-05-16 &2020-05-17 &Pearson &18271 &2020-08-12 \\
PR &2020-05-10 &2020-05-07 &Lognormal &4038 &2020-08-04 \\
SC &2020-05-16 &2020-05-20 &Pearson &15329 &2020-08-13 \\
RS &2020-05-08 &2020-05-08 &Gamma &4269 &2020-08-03 \\
MT &2020-04-30 &2020-30-04 &Pearson &701 &2020-07-30 \\
GO &2020-05-03 &2020-05-05 &Pearson &2245 &2020-08-03 \\
\bottomrule
\end{tabular}
\label{tab:tb1}
\end{table}

\section{Discussion}
\label{sec:discussion}

As already explained, the LSTM based approaches did not work well in the problem of modeling the Covid-19 dynamics. The LSTM-SAE was also tried and performed a little better. There are some explanations for this lower performance. The first issue is related to data non-linearity caused by under sampling (every other day and then every day for example) and under notifications (numbers under real values), which have also been problematic for other countries than Brazil. Several countries are under-testing their population, making the number of reported cases below reality. Even for the countries that are doing massive testing, there are often delays between the real occurrences and notification. Another potential source of error is the randomization of the weights, which can be solved with LSTM-SAE \cite{Sagheer2020}, however the first issue still remains a problem here (non-linearity problem). Yet, instability has been acknowledged during training. Several attempts had to be done in order to get a more stable model by manually tuning a fixed initialization seed.

Neural networks are known to be good function approximators, and at first look, the functions they are approximating are likely to be nonlinear. In particular, an LSTM creates an embedding that transforms the function into a linear one for the final prediction. However this is not related to the fact that the input is nonlinear, which is the case for the data distribution of Covid-19. Actually, we conjecture that the input data can be considered quasi-linear (somehow between nonlinear and linear) and that it obeys a certain pattern, otherwise no model could approximate it. The limited latent space is also a problem, even more for the long sequences as it is the case here. Besides modeling well the long-term memories, it fails in regularizing for other sequences with different properties \cite{Goodfellow-et-al-2016}. That is to say that if a certain situation (lock-down or distancing) is kept, thus it could perform better. Besides, the problem of learning long-term dependencies remains as one of the main challenges in deep learning \cite{Goodfellow-et-al-2016}. A last problem with LSTM is that the time series has to be stationary and with stable mean, an assumption that does not hold with the data that we analyzed in this paper.

\begin{figure}
    \centering
    \includegraphics[width=\textwidth]{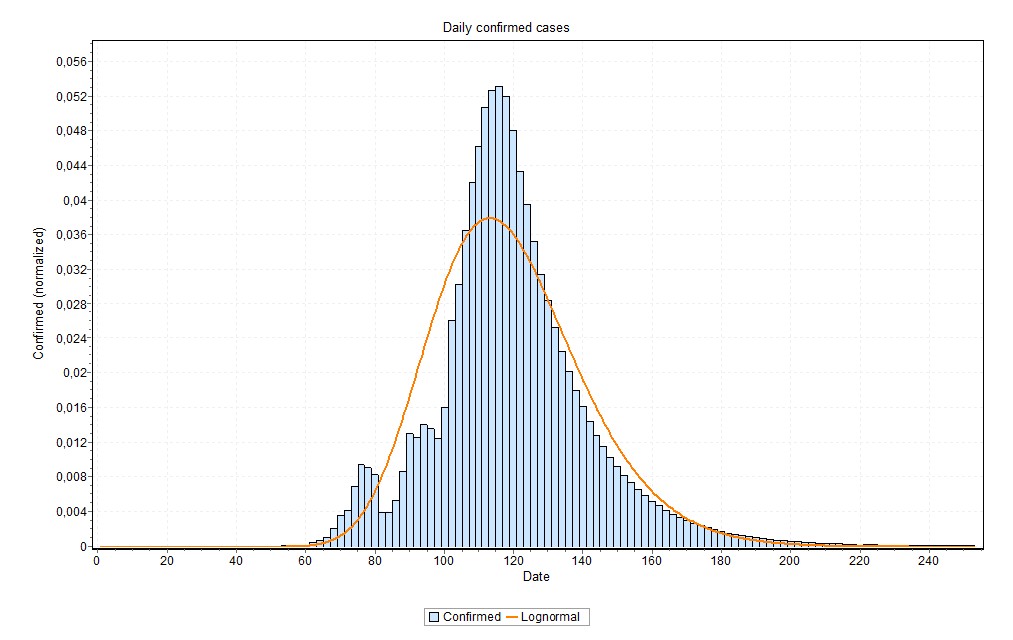}
    \caption{Curve fitting for Rio Grande do Norte state (logNormal model was the best fit) with peak indicated on May 13th, 2020.}
    \label{fig:fittingRN}
\end{figure}

An issue that recalled attention is that for states approaching the peak, the final curve fitting process performed better with a curve visually closer to data as is the cases reported in Figures \ref{fig:fittingRJ} and \ref{fig:fittingSP}. Notice in Figure \ref{fig:fittingRN} for example, this may indicate that the values close to the peak might have lower values than the ones that are predicted by MAE approach. This can be confirmed when the peak is reached. If this is the case, some adjustment can be done in our method in order to account for this property, which is our first idea for future works.

The clustering approach proposed in this paper uses a feature representation focusing on the early response of the countries. This was based on the assumption that the first week of the spread of the disease are crucial to determine its dynamics in a given region. However, in future work, it might be interesting refine the groups based on the most recent data, in order to obtain even more accurate predictions. For example, if a state is at 6 weeks after outbreak, we could compute the features for weeks 4 to 6 after outbreak.

\section{Conclusions}

The main problem that was solved in this paper is the model estimation for the Covid-19 dynamics that can be more realistic, by using cases that have already occurred in other locations or countries, with some similar distribution. Although our study focus on the Brazilian reality, technically, the proposed approach can be applied elsewhere. For determining these similar distributions, firstly, a clustering was applied to the countries/regions (training and to be predicted). This clustering was a key of the process and will be improved in future works in order to represent more closely the characteristics of the time series. 

Thus, to this end we have proposed alternative ways for modeling Covid-19 dynamics, using a data driven approach based on MAE. By our results, this approach performed better than traditional and LSTM approaches. To do that, we have proposed an initial clustering of the training data based on Early Mortality, Days until 10x, and Early Acceleration using data from regions where the pandemic is at an advanced stage. Then, we used the deep learning MAE approach to train a neural network guided by this clustering. This approach worked better, verified at the end by fitting approximating curves to the dynamics of each Brazilian state, in order to verify or ratify the peaks.

So, with basis on the results discussed above, up to date, we could verify the applicability of data driven approaches to model Covid-19 dynamics. With this approach, dealing with regional aspects based on the used features of the pandemic, city managers can get more precise information and better insight to plan their actions. Complementary material for this work can be found at \url{www.natalnet.br/covid}, where the next step is to implement this approach running and updating automatically, using the most recent available data.



\bibliography{covid}

\end{document}